\documentclass[twocolumn,showpacs,prx,amsfonts,amsmath,amssymb,floatfix]{revtex4}

\usepackage{hhline}
\usepackage{mathrsfs}
\usepackage[dvipdfmx]{graphicx}
\usepackage{dcolumn}
\usepackage{bm}% bold math
\usepackage{multirow}
\usepackage{booktabs}
\usepackage{xspace}

\newcommand{\diff}{\mathrm{d}}
\newcommand{\hhhs}{H$_3$S\xspace}
\newcommand{\hhs}{H$_2$S\xspace}
\newcommand{\tc}{$T_{\mathrm{c}}$\xspace}

\begin{document}
\title{
Effect of van Hove singularities on high-\tc superconductivity in \hhhs
}

\author{Wataru Sano$^{1,2}$}
\author{Takashi Koretsune$^{2,3}$\footnote[1]{Author to whom correspondence should be adressed: takashi.koretsune@riken.jp}}
\author{Terumasa Tadano$^{1}$}
\author{Ryosuke Akashi$^{4}$}
\author{Ryotaro Arita$^{2,5}$}

\affiliation{$^1$Department of Applied Physics, University of Tokyo,
7-3-1 Hongo, Bunkyo-ku, Tokyo 113-8656, Japan} 
\affiliation{$^2$RIKEN Center for Emergent Matter Science, 2-1 Hirosawa,
Wako, Saitama 351-0198, Japan}
\affiliation{$^3$JST, PRESTO, 4-1-8 Honcho, Kawaguchi, Saitama
332-0012, Japan}
\affiliation{$^4$Department of Physics, University of Tokyo,
7-3-1 Hongo, Bunkyo-ku, Tokyo 113-0033, Japan}
\affiliation{$^5$ERATO Isobe Degenerate π-Integration Project, Tohoku
University, Aoba-ku, Sendai 980-8578, Japan}
\date{\today}

\begin{abstract}
One of interesting open questions for the high transition temperature
 (\tc) superconductivity in sulfur hydrides is why high pressure
 phases of \hhhs have extremely high \tc's. 
Recently, it has been pointed out that the presence of the van Hove
 singularities (vHs) around the Fermi level is crucial.
However, while there have been quantitative
 estimates of \tc based on the Migdal-Eliashberg theory, 
 the energy dependence of the density of states 
 (DOS) has been neglected to simplify the Eliashberg equation.
In this study, we go beyond the constant DOS approximation and
 explicitly consider the electronic structure over 40\,eV around the
 Fermi level.
In contrast with the previous conventional calculations, this approach
 with a sufficiently large number of Matsubara frequencies enables us to
 calculate \tc without introducing the empirical pseudo Coulomb
 potential.
We show that while \hhhs has much higher \tc than \hhs for which the
 vHs is absent, the
 constant DOS approximation employed so far 
 seriously overestimates (underestimates)
 \tc by $\sim 60$\,K ($\sim 10$\,K) for \hhhs (\hhs). 
We then discuss the impact of the strong electron-phonon coupling on the
 electronic structure with and without the vHs and how it affects the
 superconductivity. 
Especially, we focus on (1) the feedback effect in the self-consistent
 calculation of the self-energy, (2) the effect of the energy shift due to
 the zero-point motion, and (3) the effect of the changes in the phonon
 frequencies 
 due to strong anharmonicity. 
We show that the effect of (1)-(3) on \tc is about $10$-$30$\,K for
 both \hhhs and \hhs.
Eventually, \tc is estimated to be $181$\,K for \hhhs at $250$\,GPa
 and $34$\,K for \hhs at $140$\,GPa,
 which explains the pressure dependence of \tc observed in the
 experiment. 
In addition, we evaluate the lowest order vertex correction beyond the
 Migdal-Eliashberg theory and discuss the validity of the Migdal approximation
 for sulfur hydrides. 

\end{abstract} 
\maketitle 

\section{Introduction} \label{sec.intro} 

%%%%%%
%
% with vertex correction
%
%%%%%%

%Realization of the superconductivity at higher temperatures has attracted
%much interest. 
Realization of superconductivity at very high temperatures has been
%a research topic with broad interest.
the Holy Grail in condensed matter physics. 
%Although cuprates %have exhibited the highest superconducting
%have held the record of the highest superconducting 
%transition temperature (\tc) for many years~\cite{cuprates_record}, 
%Although the superconducting transition temperature (\tc) of the 
%cuprate superconductors has occupied the first place in the record 
%for many years~\cite{cuprates_record}, 
%the unconventional pairing mechanism behind is yet to be fully
%understood.
While unconventional superconductors such as the cuprates~\cite{cuprates_discovery} and iron-based
superconductors~\cite{iron-based_SC_hosono} have been extensively
studied, 
the mechanism for the high transition temperatures (\tc's) is yet to 
be fully understood. 
%the unconventional pairing mechanism going beyond the standard BCS
%theory~\cite{BCS} is yet to be fully understood. 
%Therefore, it is difficult to obtain an explicit way toward higher-\tc
%superconductivity in these materials. 
%Instead, 
On the other hand, 
%there is a simple idea for high-\tc
%superconductor in 
there has been a simple but promising strategy to achieve high \tc for 
conventional phonon mediated
superconductors~\cite{ashcroft_pureH,ginzburg_pureH,ashcroft_Hrich}. 
According to the BCS theory~\cite{BCS}, 
\tc is scaled by the inverse square root of the atomic mass. 
Thus, compounds comprising light elements are promising
candidates for high-\tc superconductors.
%Such a simple prediction has been
%justified by discoveries of many superconductors, 
%such as the graphite
%intercalation compounds~\cite{gic}, 
%lithium under pressure~\cite{li_under_p,li_under_p_2,akashi_d}, %~[\onlinecite{akashi_d}], 
%magnesium diboride~\cite{mgb2}, 
%and boron-doped diamond~\cite{b-diamond_elimov,b-diamond_takano}, 
%and they exhibited relatively higher-\tc than simple metal superconductors.
Indeed, high-\tc superconductivity has been found so far in a
variety of light-element compounds 
such as the graphite intercalation compounds~\cite{gic}, 
elemental lithium under high pressures~\cite{li_under_p,li_under_p_2,akashi_d}, 
magnesium diboride~\cite{mgb2}, 
and boron-doped diamond~\cite{b-diamond_elimov,b-diamond_takano}.

Since hydrogen has the lightest atomic mass, the metallic hydrogen~\cite{ashcroft_pureH,ginzburg_pureH} or
hydrogen-rich compounds~\cite{ashcroft_Hrich} have been long 
expected to be high-\tc superconductors. 
%Recently, it has been reported that sulfur hydrides under high pressures show
%superconductivity at $190$\,K~\cite{sh3_arXiv}, and in the following
%report \tc increases up to $203$\,K~\cite{sh3_nat}. 
%This transition temperature breaks 
%%a record of \tc which cuprates
%%had~\cite{cuprates_record}. 
%the record of \tc of cuprates: $130$\,K at ambient
%pressure~\cite{cuprates_ambient}, $164$\,K
%at high pressure~\cite{cuprates_record}.
Recently, it has been reported that \hhs under pressures of
$100$-$200$\,GPa  exhibits 
superconductivity at extremely high temperatures up to $\sim 200$\,K~\cite{sh3_arXiv, sh3_nat}, 
breaking the record of the cuprates~\cite{cuprates_record, cuprates_ambient}.
%Due to the observation of the isotope effect, its superconductivity
%should be related to the atomic vibration. 
%In addition, two phases were observed and these phases had different
%\tc. 
%In the relatively low-\tc phase,
%\tc ranges from $50$\,K to $150$\,K, while it reaches about $200$\,K 
%in the high-\tc phase~\cite{sh3_nat,sh3_shimizu}.

Prior and after this experimental discovery, there are
a lot of {\it ab initio} studies for compressed sulfur
hydrides~\cite{ma_jpc,duan_SciRep,li_arxiv,hirsch_physica,mazin_prb,duan_prb_15,
pickett_prb,gross_arxiv,akashi_prb,mauri_prl,arxiv_bianconi,ma_arxiv_15,
pickett_arxiv,boeri_prb,yao_arxiv,pietronero_arxiv_15_1,pietronero_arxiv_15_2,gorkov_arxiv,mauri_arxiv_15,akashi_hxs}. 
%% to arita Sci.Rep.は particle swarm でなく generic algorithm
%Based on structure searching calculations, 
%it is proposed that the low-\tc phase has \hhs composition, 
%whereas \hhhs is realized at
%the high-\tc phase~\cite{duan_SciRep,mazin_prb,duan_prb_15,gross_arxiv,akashi_prb}. 
A variety of possible crystal structures have been found by structure searching 
calculations~\cite{ma_jpc,duan_SciRep,duan_prb_15,mauri_arxiv_15,akashi_hxs}, 
and \tc has been estimated to be lower than $100$\,K (as high as $200$\,K) 
for \hhs (\hhhs)~\cite{duan_SciRep,mazin_prb,gross_arxiv,akashi_prb,mauri_prl,ma_arxiv_15,pietronero_arxiv_15_2,mauri_arxiv_15}. 
It is also suggested that chemical substitution of sulfur atoms
could enhance \tc~\cite{boeri_prb,yao_arxiv}.
Most of these works have concluded that the compressed sulfur hydrides are 
phonon-mediated strong-coupling superconductors. 
%strong-coupling superconductors induced by the conventional phonon
%mediated mechanism. 
Not only the existence of high frequency phonons due to the
hydrogen motion,
strong electron-phonon coupling has also been shown to be %important factor for the realization
%of high-\tc as conventional superconductors, 
%especially in the \hhhs phases.
important for high-\tc superconductivity, especially in \hhhs. 
%As for the pairing mechanism, the atomic vibration is expected to play a key role, 
%since the isotope effect is
%significant~\cite{sh3_arXiv,sh3_nat,sh3_shimizu}.
These results are indeed consistent with the experiment~\cite{sh3_arXiv,
sh3_nat, sh3_shimizu} where the isotope effect is observed to be
significant. 

Interestingly, in the calculations based on the Migdal-Eliashberg (ME)
theory~\cite{migdal,eliashberg}, 
there is a clear difference in calculated \tc's between 
\hhs and \hhhs. 
While the origin of this difference is yet to be fully
understood, 
recently, it has been suggested that van Hove singularities (vHs) in the electronic
structure of \hhhs 
play a key role to understand this problem~\cite{pickett_prb, arxiv_bianconi,pickett_arxiv,pietronero_arxiv_15_1}.
%To our knowledge, 
%the effects of the van Hove singularities have not been estimated
%quantitatively in the calculations of \tc based ob the
%ME theory~\cite{pickett_prb,
%arxiv_bianconi,pickett_arxiv}. 
It is noteworthy that this situation is similar to that of the A15
compounds, 
for which the density of states (DOS) has a sharp peak around the Fermi
level~\cite{carbotte_ssc_78,pickett_prb_82,pickett_prl_82,carbotte_cjp_83,carbotte_prb_83}. 
Indeed, there have been some model calculations which studied how the
energy dependence of the DOS affects \tc~\cite{pickett_prb_82}.
On the other hand, for \hhhs, the effect of the vHs
on the superconductivity has not been fully understood. 
In the previous studies based on the ME
theory~\cite{ma_jpc,duan_SciRep,li_arxiv,pietronero_arxiv_15_2},
%the constant DOS approximation has been employed, in which 
the DOS is assumed to have no energy dependence. 
%That is because energy dependence of density of states (DOS) is usually 
%neglected to simplify the Eliashberg equation. 
This is mainly because one can reduce the numerical cost to solve the
Eliashberg equation. 
%In this study, we seriously consider how structures of DOS affect
%\tc of sulfur hydrides within the framework of the ME theory. 
%In this study, we examine how the presence/absence of the vHs
%affects \tc of sulfur hydrides within the framework of
%the ME theory.  
%The ratio between \tc and the band width is usually extremely small
%(less than $\mathcal{O}(10^{-5})$), so that the number of Matsubara
%frequencies to cover the energy range of the band width in the
%Eliashberg equation becomes formidably huge. 
%Thus we introduce a cutoff for the Matsubara frequency and the empirical
%pseudo Coulomb potential, $\mu^*$, which represents the retardation
%effect. 
%On the other hand, for sulfur hydrides, the ratio between \tc and the
%band width is only $\mathcal{O}(10^{-3})$. 
%In this study, we have performed calculation where we explicitly take
%account of the energy dependence of the DOS. 
%By considering the retardation effect directly, we succeeded in
%evaluating \tc without introducing any adjustable parameters such as
%$\mu^*$. 

In this study, we examine how the presence/absence of the vHs affects
\tc of sulfur hydrides. 
To this end, we go beyond the constant DOS approximation. 
With a sufficiently large number of Matsubara frequencies, the
retardation effect is automatically considered. 
Such a calculation is possible because the ratio between \tc and the
band width is only $\mathcal{O}(10^{-3})$. 
Note that the ratio for usual conventional superconductors is as small
as $\mathcal{O}(10^{-5})$ so that the retardation effect is represented
by introducing the empirical pseudo Coulomb potential $\mu^{*}$.

%In order to discuss the effect of the van Hove singularities, 
%the normal self-energy should be accurately evaluated by including the
%strong coupling between electrons and lattices. 
%However, in the conventional calculations, the feedback effect in the
%normal self-energy is absent since self-consistency between the
%normal Green's function and self-energy does not retained. 
%In our calculation, such self-consistency is also taken into account. 
Another advantage of the present approach is that we can calculate the
self-energy due to the electron-phonon coupling self-consistently. 
As will be discussed in Sec.\,\ref{sec.method}, in the constant DOS
approximation, %only the one-shot self-energy can be obtained. 
the feedback effect included in self-consistent calculation is 
automatically neglected.
Also in the standard density functional theory for
superconductors~\cite{gross_arxiv,akashi_prb,gross_prb_05}, the exchange
correlation functional
representing the mass enhancement effect is not calculated
self-consistently. 
In this study, we discuss how the self-consistency in the calculation of
the self-energy affects the superconductivity in the sulfur hydride
superconductors.

%In systems with light elements, large amplitudes of the zero-point motion are
%expected. 
%The zero-point motion causes anther modulation of the self-energy, which
%is the so called zero-point renormalization (ZPR).
%Since the sulfur hydrides include the hydrogen, the ZPR might not be
%negligible. 
%As for the self-energy due to the electron-phonon coupling, 
In the calculation of the self-energy, 
we first consider
the standard contribution (the lowest phonon-exchange diagram) in the
ME theory. 
On top of that, we then study the effect of the so-called zero-point
renormalization (ZPR), i.e., the band energy shift due to the zero-point
motion. 
It should be noted that the amplitude of the zero-point motion of
hydrogen atoms in sulfur hydrides is lager than 0.1\,\AA.
Indeed, it has been proposed that its effect on the electronic structure
and superconductivity is expected to be
significant~\cite{arxiv_bianconi} 
and contribute to the stability of high-symmetry cubic 
phase~\cite{mauri_arxiv_15}.

%For calculations of superconducting pairing interaction and the
%self-energy, one should evaluate the phonon frequency and
%electron-phonon coupling constant. 
%In order to calculate phonon properties, the density functional
%perturbation theory~\cite{dfpt} is employed, usually with the harmonic
%phonon approximation. 
Another characteristic feature of the superconducting sulfur hydrides is
its strong anharmonic effect. 
Recently, within the constant DOS approximation, it has been shown that
the anharmonicity in \hhhs significantly suppresses the
superconductivity, especially when the system experiences the second
order structural phase transition (from $R3m$ to
$Im\bar{3}m$)~\cite{mauri_prl}. 
In this work, we also study the anharmonic effect in the
energy-dependent Eliashberg approach for \hhhs and \hhs, and show that
the impact of anharmonicity is significant not only in \hhhs but also
\hhs.

Finally, we study the validity of the ME theory. 
It has been suggested that the Migdal theorem~\cite{migdal} might not be applicable to
\hhhs ~\cite{arxiv_bianconi,gorkov_arxiv},
%due to small Fermi pockets, some of which have contributions to the vHs~\cite{arxiv_bianconi,gorkov_arxiv},
though \tc estimated by the ME theory is
consistent with experimentally observed values.
Since the effective Fermi energy at the vHs is small and
comparable to the phonon energy scale, the premise of the ME theory
might not be satisfied. 
%On the other hand, 
%since \tc estimated by the ME is
%consistent with experimentally observed values,
%the ME theory looks working well in sulfur hydrides
Indeed, there has been a study proposing unconventional pairing
mechanism~\cite{hirsch_physica}. 
To answer this question, we estimate the lowest order vertex
correction beyond the ME theory and examine its effect on \tc.

This paper is organized as follows.
In Sec.\,\ref{sec.method} we review the methods to study the normal and
superconducting properties of solids from first-principles.
%In particular, we take special care of the variety of the linearized
%Migdal-Eliashberg theory which evaluate \tc based on the different
%approximations.
In particular, we describe the 
approximations employed thus far to solve the linearized Eliashberg
equation and discuss how we go beyond. 
%This is important 
%since the \tc
%sensitively depends on the approximation employed. 
%Here we also summarize the linear response theory and direct method for
%phonons to treat the ZPR and anharmonicity.
Here we also describe how to treat the ZPR and anharmonicity. 
In Sec.\,\ref{sec.elph}, we discuss the normal electronic structure and the
phonon structure for \hhhs with the $Im\bar{3}m$ structure and \hhs with
the $P\bar{1}$ structure. 
%of sulfur hydrides for two representative phases. %several phases. 
In Sec.\,\ref{sec.eliash},  
%we investigate the validity of the 
%methods introduced in Sec.\,\ref{sec.method}. %through the evaluation 
%\tc for sulfur hydrates.
%In addition, we study how the ZPR affects \tc. %in sulfur hydrides.
%We calculate \tc considering both the ZPR and anharmonicity
%with energy dependent DOS and discuss which has significant
%contribution to \tc in Sec.\,\ref{sec.anharm}.
we show that the constant DOS approximation seriously overestimates
\tc of \hhhs by $\sim 60$\,K. 
On the other hand, in the case of \hhs for which the vHs
are absent, the constant DOS approximation underestimates
\tc by $\sim 10$\,K. 
We then discuss how the self-energy due to the strong electron-phonon
coupling affects the van Hove singularities and \tc. 
We study (1) the feedback effect in the self-consistent calculation of
the self-energy, (2) the effect of the electron energy shift due to the
zero-point motion, and (3) the effect of the changes in the phonon
frequencies due to the strong anharmonicity. 
We show that the effect of (1)-(3) on \tc is about $10$-$30$\,K for
both \hhhs and \hhs, and \tc is estimated to be $181$\,K for
\hhhs, and $34$\,K for \hhs. 
These results suggest that \hhhs (\hhs) is responsible for the high-
(low-)\tc superconductivity under pressures higher (lower) than $\sim
150$\,GPa. 
In Sec.\,\ref{sec.vertex}, we evaluate the lowest order vertex
correction and its effect on \tc in order to obtain the criterion for the
justification of the ME theory.
Finally, we give a summary of this study in Sec.\,\ref{sec.summary}.
\section{Method} \label{sec.method}

%%%%%%%%%%%%%%%%%%%%%%%%%%%%%%
\subsection{Migdal-Eliashberg theory for \tc calculation with energy dependent DOS}
%We perform the fully non-empirical \tc calculation based on the
%Migdal-Eliashberg (ME) theory~\cite{migdal,eliashberg}. 
Based on density functional and density functional perturbation
theory (DFPT)~\cite{dfpt}, one can obtain the following Hamiltonian for electron-phonon
coupled systems:
\begin{align}
 \label{eq.Hamiltonian}
 H_{\rm ep}=H_{0} + H_{\rm el\mathchar`-el} + H_{\rm el\mathchar`-ph},
%&\sum_{n, \bm{p},\sigma} 
%  \xi_{n\bm{p}} c_{n\bm{p}\sigma}^{\dagger} c_{n \bm{p} \sigma}
%  + \sum_{\bm{q},\lambda} \omega_{\bm{q} \lambda}
% b_{\bm{q}\lambda}^{\dagger}  b_{\bm{q}\lambda} \nonumber \\
%%
% & + \frac{1}{2N} \sum_{\bm{q} \neq 0} \sum_{nmkl, \bm{p}\bm{p}',\sigma
%  \sigma ' } V_{c\bm{p},\bm{p}'}^{nm,kl} (\bm{q})
%  c_{n\bm{p} + \bm{q}\sigma}^{\dagger} c_{m\bm{p}' -
%  \bm{q}\sigma'}^{\dagger} c_{l\bm{p}' \sigma'} c_{k\bm{p} \sigma}
%  \nonumber\\
%%
%%
%  &+ \frac{1}{\sqrt{N}}\sum_{\bm{q} \neq 0, \lambda} \sum_{nm,\bm{p},\sigma}
%  g^{n{\bm p} + {\bm q}, m {\bm p} }_{\lambda} ({\bm q})
%  (b_{\bm{q}\lambda} +  b_{-\bm{q}\lambda}^{\dagger} ) c_{n\bm{p} + \bm{q}\sigma}^{\dagger} c_{m\bm{p} \sigma}
\end{align}
where 
\begin{align}
 H_{0} = \sum_{j, \bm{p},\sigma} 
  \xi_{j\bm{p}} c_{j\bm{p}\sigma}^{\dagger} c_{j \bm{p} \sigma}
  + \sum_{\bm{q},\lambda} \omega_{\bm{q} \lambda}
 b_{\bm{q}\lambda}^{\dagger}  b_{\bm{q}\lambda}, 
\end{align}
\begin{align}
 H_{\rm el\mathchar`-el} = 
 \frac{1}{N} \sum_{\bm{q} \neq 0} \sum_{jl, \bm{p}}
 V^{\mathrm{c}} (\bm{q})
 c_{j\bm{p}+\bm{q} \uparrow}^{\dagger} c_{j -\bm{p}-\bm{q} 
  \downarrow}^{\dagger} c_{l -\bm{p} \downarrow} c_{l\bm{p} \uparrow}, 
\end{align}
\begin{align}
 H_{\rm el\mathchar`-ph} =
  \frac{1}{\sqrt{N}}\sum_{\bm{q} \neq 0, \lambda} \sum_{jl,\bm{p},\sigma}
  g^{j{\bm p} + {\bm q}, l {\bm p} }_{\lambda} ({\bm q}) 
  &(b_{\bm{q}\lambda} +  b_{-\bm{q}\lambda}^{\dagger} ) \nonumber \\
 & \times c_{j\bm{p} + \bm{q}\sigma}^{\dagger} c_{l\bm{p} \sigma},
\end{align}
with $c_{j\bm{p}\sigma}^{\dagger}$ ($c_{j\bm{p}\sigma}$) being a
creation (annihilation) operator of an electron with spin $\sigma$ and
momentum $\bm{p}$ in the $j$-th band, $\xi_{j \bm{p}}$ being an electron 
dispersion with respect to the Fermi level, 
$ b_{\bm{q}\lambda}^{\dagger}$ ($b_{\bm{q}\lambda}$)
being a creation (annihilation) operator of a phonon with momentum
$\bm{q}$ and mode $\lambda$, 
$\omega_{\bm{q} \lambda}$ being a phonon frequency, 
$g^{j{\bm p} + {\bm q}, l {\bm p} }_{\lambda} ({\bm q})$ being a
electron-phonon matrix element defined by Eq.\,\eqref{eq.elph_matrix}, 
and $V^{\mathrm{c}}$ being the bare
electron-electron Coulomb interaction.
Here we consider the Coulomb repulsion between electrons only for
pairing channels explicitly since electron dispersion $\xi_{j\bm{p}}$
already includes the contribution of the direct and exchange channel
of the Coulomb interaction at the mean-field level. 
Such a treatment is justified for weakly correlated systems like conventional
superconductors.

The problem for conventional superconductivity is how to treat the
Hamiltonian given by Eq.\,(\ref{eq.Hamiltonian}).
Fortunately the Migdal theorem greatly simplifies complicated many-body
problem of the electron-phonon coupled system through neglecting the
vertex correction~\cite{migdal}. 
Within the framework of the Migdal-Eliashberg 
theory~\cite{migdal,eliashberg}, the self-energy is given by
\begin{align}
  \label{eq.eliash1}
  \Sigma_{j\bm{p}}(i\omega_n) = %\nonumber  \\
 -\frac{1}{ N \beta} &\sum_{l\bm{q}m}  % \nonumber \\
 \tilde{V}^{\mathrm{ph}}_{j\bm{p}+\bm{q}, l\bm{p}} (\bm{q}, i\omega_m) \nonumber \\
 &\times G_{l\bm{p+q}} (i\omega_m + i\omega_n),
\end{align}
\begin{align}
 \label{eq.eliash2}
  \Delta_{j\bm{p}}(i\omega_n) = \frac{1}{N\beta} 
 &\sum_{l {\bm q} m}
  \{\tilde{V}^{\mathrm{ph}}_{j\bm{p}+\bm{q}, l\bm{p}} (\bm{q}, i\omega_m) +
  \tilde{V}^{\mathrm{c}}_{j\bm{p}+\bm{q}, l\bm{p}} (\bm{q},
  i\omega_m) \} \nonumber \\
 &\times F_{l\bm{p+q}} (i\omega_m + i\omega_n),
\end{align}
where $j$ and $l$ are the band indices, $\Sigma_{j\bm{p}}(i\omega_n)$ and
$\Delta_{j\bm{p}}(i\omega_n)$ are
the normal and the anomalous self-energies, and $G_{j\bm{p}} (i\omega_n)$ and
$F_{j\bm{p}} (i\omega_n)$ are the electron normal and anomalous Green's functions.
Here the band off-diagonal elements of the Green's function are
neglected. 
If the density functional calculation is a good starting point for
superconductors, the off-diagonal elements can be safely ignored.
 $\tilde{V}^{\mathrm{ph}}$ 
is the screened electron-electron interaction mediated by phonons. It is given by 
\begin{eqnarray}
 \label{eq.screened_ph}
  \tilde{V}^{\mathrm{ph}}_{j\bm{p}+\bm{q}, l\bm{p}} (\bm{q}, i\omega_m)
  = \sum_{\lambda} | \tilde{g}^{j\bm{p+q}, l\bm{p}}_\lambda (\bm{q}) |^2
  D_{\bm{q} \lambda} (i\omega_m),
\end{eqnarray}
where $\tilde{g}^{j\bm{p+q}, l\bm{p}}_\lambda (\bm{q})$ and $D_{\bm{q} \lambda} (i\omega_m)$
denote the screened electron-phonon matrix element and the phonon Green's
function. Here $\tilde{g}$ and $D$ are
%renormalized quantities 
screened quantities 
and should include the static screening
effect by the electron polarization. 
On the other hand, in {\it ab initio} calculations based on density functional theory, 
calculated $g$ and the phonon frequency already include such
screening effects in the static level.
Therefore, one can consider $\tilde{g}$ in Eq.\,(\ref{eq.screened_ph})
as $g$ from DFPT and $D$ as the free phonon Green's
function defined by
\begin{align}
 D_{\bm{q}\lambda} (i\omega_m) = -
 \frac{2\omega_{\bm{q}\lambda}}{\omega_m^2 + \omega_{\bm{q}\lambda}^2},
\end{align}
where $\omega_{\bm{q}\lambda}$ is also calculated by DFPT. 
%Hereafter, we identify $\tilde{g}$ with $g$ and $D$
%with the free phonon Green's function.

%$\tilde{V}^{c}_{j\bm{p}, l\bm{p}'} (i\omega)
% = \langle \psi_{j\bm{p}\uparrow} \psi_{j-\bm{p}\downarrow} |
% \epsilon^{-1}(i\omega) V^{c}| \psi_{l\bm{p}'\uparrow} \psi_{l-\bm{p}'\downarrow} \rangle$ 
%is the screened Coulomb interaction for the paring channel.
% The screened Coulomb interaction is calculated thorough the symmetrized dielectric function~\cite{symm_diele}
%\begin{align}
%  \tilde{\epsilon}_{\bm{G} \bm{G}'}(\bm{q}; i\omega) =
%  \delta_{\bm{G} \bm{G}'} -
%  \frac{4\pi}{\Omega} 
%%
%  \frac{1}{|\bm{q}+\bm{G}|} 
%%
%  \chi_{\bm{G} \bm{G}'}(\bm{q}; i\omega)
%%
%  \frac{1}{|\bm{q}+\bm{G}'|},
%\end{align}
%where $\chi$ is the polarization function, $\bm{G}$ is the reciprocal
%lattice vector, and $\Omega$ is the volume of the unit cell.
%With the symmetrized dielecrtic function, the screened Coulomb
%interaction is given by
%\begin{align}
% \tilde{V}^c_{j\bm{p}+\bm{q}, l\bm{p}}(i\omega) =
%  \frac{4\pi}{\Omega}  \sum_{\bm{G}\bm{G}'}
%  \frac{ \rho^{j \bm{p} + \bm{q} }_{l \bm{p}} (\bm{G}) 
%%
%  \tilde{\epsilon}^{-1}_{\bm{G} \bm{G}'}(\bm{q}; i\omega)
%%
%  \{\rho^{j \bm{p} + \bm{q} }_{l \bm{p}} (\bm{G}')\}^* }{
%%%
%  |\bm{q}+\bm{G}||\bm{q}+\bm{G}'|}
%\end{align}
%with
%\begin{align}
%  \rho^{j \bm{p} + \bm{q}}_{l \bm{p}} (\bm{G}) = 
%  \int_{\Omega} \diff^3 r \psi_{j \bm{p} + \bm{q}}^* (\bm{r})
%  e^{i ( \bm{q} + \bm{G}) \cdot \bm{r}} 
%  \psi_{l \bm{p} } (\bm{r}),
%\end{align}
%where $\int_{\Omega}$ denotes the integration in the unit cell. 
The screened Coulomb interaction for the pairing channel, 
$\tilde{V}^{\mathrm{c}}_{j\bm{p}, l\bm{p}'} (i\omega_m)
 = \langle \psi_{j\bm{p}\uparrow} \psi_{j-\bm{p}\downarrow} |
 \epsilon^{-1}(i\omega_m) V^{\mathrm{c}}| \psi_{l\bm{p}'\uparrow} \psi_{l-\bm{p}'\downarrow} \rangle$ 
is calculated through the symmetrized dielectric function~\cite{symm_diele},
$\tilde{\epsilon}_{\bm{G} \bm{G}'}$, as
\begin{align}
 \label{eq.rpa_coulomb}
 \tilde{V}^{\mathrm{c}}_{j\bm{p}+\bm{q}, l\bm{p}}(i\omega_m) =
  \frac{4\pi}{\Omega}  \sum_{\bm{G}\bm{G}'}
  \frac{ \rho^{j \bm{p} + \bm{q} }_{l \bm{p}} (\bm{G}) 
  \tilde{\epsilon}^{-1}_{\bm{G} \bm{G}'}(\bm{q}; i\omega_m)
  \{\rho^{j \bm{p} + \bm{q} }_{l \bm{p}} (\bm{G}')\}^* }{
  |\bm{q}+\bm{G}||\bm{q}+\bm{G}'|},
\end{align}
\begin{align}
  \tilde{\epsilon}_{\bm{G} \bm{G}'}(\bm{q}; i\omega_m) =
  \delta_{\bm{G} \bm{G}'} -
  \frac{4\pi}{\Omega} 
  \frac{1}{|\bm{q}+\bm{G}|} 
  \chi_{\bm{G} \bm{G}'}(\bm{q}; i\omega_m)
  \frac{1}{|\bm{q}+\bm{G}'|}.
\end{align}
Here, $\chi$ is the polarization function, $\bm{G}$ is the reciprocal
lattice vector, $\Omega$ is the volume of the unit cell and $\rho^{j
\bm{p} + \bm{q}}_{l \bm{p}} (\bm{G})$ is written as
\begin{align}
  \rho^{j \bm{p} + \bm{q}}_{l \bm{p}} (\bm{G}) = 
  \int_{\Omega} \diff^3 r \psi_{j \bm{p} + \bm{q}}^* (\bm{r})
  e^{i ( \bm{q} + \bm{G}) \cdot \bm{r}} 
  \psi_{l \bm{p} } (\bm{r}),
\end{align}
where $\int_{\Omega}$ denotes the integration in the unit cell.

In this study, the screening is treated within the random phase
approximation (RPA)~\cite{pines}. In the RPA, the polarization function
is given by the following equation:
\begin{align}
  \chi_{\bm{G} \bm{G}'}(\bm{q}; i\omega_m) = \frac{2}{\Omega} 
  \sum_{\bm{p}} \sum_{j:\rm{unocc}, l:\rm{occ}} 
  [\rho^{j \bm{p} + \bm{q}}_{l \bm{p}} (\bm{G}) ]^*
  \rho^{j \bm{p} + \bm{q}}_{l \bm{p}} (\bm{G}') \nonumber \\
  \times \biggl\{ 
   \frac{1}{i\omega_m - \xi_{j \bm{p} + \bm{q}} +
   \xi_{l \bm{p}}}
   - \frac{1}{i\omega_m + \xi_{j \bm{p} + \bm{q}} -
   \xi_{l \bm{p}}}
   \biggr\}.
\end{align}
Here the polarization function depends on the Matsubara frequency. 
However, we
ignore this frequency dependence and treat the screened Coulomb
interaction as a static repulsion between the paired electrons. One should
notice that the structure of the Coulomb interaction along the frequency
direction leads to an enhancement of \tc by the plasmon mechanism~\cite{takada_plasmon}. 
The inclusion of the frequency dependence for the screened Coulomb
interaction is left as a future work.

For the calculation of \tc, the second order
products of the anomalous quantities can be ignored. Therefore, the 
equations are
linearized and the anomalous Green's function is reduced to the product
of the normal Green's function and the anomalous self-energy like
$F_{j\bm{p}} (i\omega_n)= - G_{j\bm{p}} (i\omega_n)G_{j-\bm{p}}
(-i\omega_n) \Delta_{j\bm{p}}(i\omega_n)$.
% Here we have suppressed the band
%index for simplicity. Inclusion of band index is straightforward.

In conventional calculations of \tc based on the ME theory, several approximations
are introduced to simplify Eqs.\,(\ref{eq.eliash1}) and~(\ref{eq.eliash2})~\cite{rainer_tc,allen_tc,carbotte_tc}.
%If the DOS could be considered as a constant around the Fermi level,
%one can completely omit the momentum sum by taking the isotropic
%limit. 
Since the pairing interaction works only for low energy states, 
we rewrite the momentum sum as an energy integral assuming 
that DOS is constant around the Fermi level. 
For the Matsubara frequency sum, one introduces a cutoff
frequency (of the order of the phonon energy scale) by replacing the Coulomb
interaction $\tilde{V}^{\mathrm{c}}$ with the pseudo Coulomb potential
$\mu^*$~\cite{mu_star}. 
%$\mu^*$ represents the retardation effect of the bare
%Coulomb interaction $\mu$ and it takes smaller values than $\mu$. 
$\mu^*$ represents the retardation effect defined by 
\begin{align}
 \label{eq.mustar_coulomb}
 \mu^*  = \frac{\mu}{ 1 + \mu \ln (\omega_{\rm el}/\omega_{\rm c})}.
\end{align}
%where $\mu$ is the unrenormalized Coulomb potential written as
Here, $\omega_{\rm el}$ and $\omega_{\rm c}$ are the cutoff frequencies
of the order of the electron and phonon energy scale, respectively, and
$\mu$ is the unrenormalized Coulomb potential written as 
\begin{align}
 \label{eq.mu_coulomb}
  N(0)\mu = \frac{1}{N^2} \sum_{jl,\bm{p}\bm{q}} 
 \tilde{V}^{\mathrm{c}}_{j\bm{p}+\bm{q}, l\bm{p}}(0)
 \delta(\xi_{j \bm{p}+\bm{q}}) \delta(\xi_{l \bm{p}}),
\end{align}
where $N(0)$ denotes the DOS at the Fermi level. 
With these approximations, the linearized version of Eqs.\,(\ref{eq.eliash1}) and\,(\ref{eq.eliash2})
are reduced to 
\begin{eqnarray}
 \label{eq.Z_iso}
  Z(i\omega_n) = 1 + \frac{1}{\omega_n}\frac{\pi}{\beta} \sum_{n'}'
  \lambda(i\omega_n - i\omega_{n'})\mathrm{sgn}(\omega_{n'}),
\end{eqnarray}%
\begin{align}
 \label{eq.gap_iso}
  \phi(i\omega_n)
  = \frac{1}{Z(i\omega_n)} \frac{\pi}{\beta}
  \sum_{n'}'\frac{\phi(i\omega_{n'})}{|\omega_{n'}|}
  \biggl\{\lambda(i\omega_n - i\omega_{n'})    
  - \mu^{*} \biggr\},
\end{align}
where $Z(i\omega_n)$ and $\phi(i\omega_n)$ denote the renormalization
function and the gap function. $\Sigma$ with prime denotes the summation
with the frequency cutoff $\omega_{\rm c}$. 
$\lambda(i\omega_m)$ is the electron-phonon coupling
\begin{eqnarray}
 \label{eq.elph}
 \lambda(z)=
  \int_{0}^{\infty} \diff \nu
  \frac{2\nu}{\nu^2 - z^2} \alpha^2 F (\nu),
\end{eqnarray}
and $\alpha^2 F (\nu)$ is the Eliashberg function defined by
\begin{align}
 \label{eq.eliash_function}
 \alpha^2 F (\nu) &= \frac{1}{N(0)} \sum_{jl, \bm{p} \bm{q}, \lambda }
  | g^{j\bm{p}+\bm{q},l\bm{q}}_\lambda (\bm{q}) |^2 \nonumber \\
 &\times
  \delta(\xi_{j\bm{p}+\bm{q}}) \delta(\xi_{l\bm{p}}) \delta(\nu -
  \omega_{\bm{q}\lambda}).
\end{align}
%where $N(0)$ and 
%$\xi_{j\bm{p}}$ denote the DOS at the Fermi
%level and the electron band energy measured from the Fermi level.
%where $N(0)$ denotes the DOS at the Fermi level. 
% and all these quantities can be calculated with DFT 
The Eliashberg function plays a central role in the conventional ME theory. 
If one knows $\alpha^2 F$, \tc can be calculated with
Eqs.\,(\ref{eq.Z_iso}) and~(\ref{eq.gap_iso}) easily. 
%In conventional calculations, $\mu^*$ is treated as an adjustable
%parameter~\cite{rainer_tc,allen_tc,carbotte_tc}. 
In conventional calculations, $\mu^*$ is not evaluated with
Eqs.\,\eqref{eq.mustar_coulomb} and~\eqref{eq.mu_coulomb} but rather
treated as an adjustable
parameter~\cite{rainer_tc,allen_tc,carbotte_tc}. 
%Hereafter we call this
%treatment constant DOS ME theory.
%Simplicity of the constant DOS ME theory is also
%related to the absence of the level shift function defined as the real
%part of the self-energy. 
%%The level shift function represents the shift of energy level. 
%Since the constant DOS approximation implies
%the particle-hole symmetry, the level shift function becomes zero at the
%Fermi level with this approximation.
%Due to the absence of the level shift function, the normal Green's
%function (which only depends on the renormalization function $Z$) is
%calculated by an one-shot way.

%In the calculation with the constant DOS approximation, 
%the pseudo Coulomb potential $\mu^*$ depends on the cutoff $\omega_c$ 
%and $\mu^* \rightarrow \mu$ in the limit of large $\omega_c$. 
%Thus, we expect that Eq.\,\eqref{eq.gap_iso} with $\mu^* = \mu$ gives
%converged \tc for sufficiently large $\omega_c$. 
%However, one hardly ever reach the convergence with fixed $\mu$ due to
%the logarithmically divergent behavior of the summation $\sum_{n}
%1/(2n+1)$
%in the r.h.s. of the gap equation. 
%Instead, one can calculate \tc with fixed $\mu$ by introducing another
%cutoff frequency $\omega_{\rm el}$ of order of the electron energy
%scale.

%Generalization to the case of rapidly varying DOS is analytically easy,
%although it needs more computational costs. 
While analytical formulation of the scheme considering 
the energy dependence of DOS is rather straightforward,
actual calculation is numerically expensive.
To mitigate the
computational costs, we take momentum average of $\tilde{V}^{\mathrm{ph}}$ and 
$\tilde{V}^{\mathrm{c}}$ like 
$\tilde{V}^{\mathrm{ph}}_{jl} (\bm{q},i\omega_m) 
= \langle \tilde{V}^{\mathrm{ph}}_{j\bm{p}+\bm{q},l\bm{p}} (\bm{q}, i\omega_m) \rangle_{\bm{p}}$
and
$\tilde{V}^{\mathrm{c}}_{jl} (\bm{q},i\omega_m) 
= \langle \tilde{V}^{\mathrm{c}}_{j\bm{p}+\bm{q},l\bm{p}} (\bm{q}, i\omega_m) \rangle_{\bm{p}}$.
In conventional superconductivity, this simplification could be a good
approximation since the gap function is almost isotropic and the complex momentum
dependence is not important. For the phonon mediated interaction, this
average is achieved by averaging the electron-phonon matrix element 
$|g^{jl}_{\lambda} ({\bm q})|^2 = \langle
|g^{j\bm{p}+\bm{q},l\bm{p}}_{\lambda} ({\bm q})|^2 \rangle_{\bm{p}}$.
The averaged interaction is given by
\begin{align}
  \tilde{V}^{\mathrm{ph}}_{jl} (\bm{q}, i\omega_m) = 
  \sum_{\lambda}  
  |g^{jl}_{\lambda}(\bm{q})|^2 
  D_{\bm{q}\lambda}(i\omega_m).
\end{align}
Then, the linearized equations are written as
\begin{align}
 \label{eq.eliash_sigma}
   \Sigma_{j\bm{p}}(i\omega_n) = 
 -\frac{1}{ N \beta} &\sum_{l\bm{q}m}
 %|\tilde{g}^{jl}_{\lambda} ({\bm q}) |^2 
 %D_{\lambda {\bm q}}(i\omega_m) \nonumber \\
 \tilde{V}^{\mathrm{ph}}_{jl} (\bm{q}, i\omega_m) \nonumber \\
&\times G_{l\bm{p+q}} (i\omega_m + i\omega_n),
 \end{align}
\begin{align}
 \label{eq.eliash_delta}
   \Delta_{j\bm{p}}(i\omega_n)& = - \frac{1}{N\beta} 
 \sum_{l {\bm q} m}
  \{\tilde{V}^{\mathrm{ph}}_{jl} (\bm{q}, i\omega_m) + \tilde{V}^{\mathrm{c}}_{jl} (\bm{q},
  i\omega_m) \} \nonumber \\
  &\times G_{l\bm{p} + \bm{q}}(i\omega_n + i\omega_m) 
  G_{l-\bm{p} - \bm{q}}(-i\omega_n - i\omega_m)  \nonumber \\
  &\times \Delta_{l\bm{p} + \bm{q}}(i\omega_n + i\omega_m).
\end{align}
Based on this formulation, one can include
the effect of energy dependent DOS on \tc.

Eq.\,(\ref{eq.eliash_sigma}) is solved with the Dyson equation:
\begin{align}
 \label{eq.g_dyson}
    G_{j\bm{p}}(i\omega_n) = \frac{1}{i \omega_n - \xi_{j\bm p} -
  \Sigma_{j\bm{p}}(i\omega_n)},
\end{align}
by either the self-consistent (SC) or one-shot way. 
%This is clear
%difference between the constant DOS ME theory and
%Eqs.\,(\ref{eq.eliash_sigma}) and\,(\ref{eq.eliash_delta}). 
%In the constant DOS ME theory, 
It should be noted that once we employ the constant DOS approximation,
one cannot perform the SC calculation for
the normal part. 
As mentioned above, this fact is ascribed to the neglect of the level
shift function in the constant DOS approximation.
On the other hand, in Eqs.\,(\ref{eq.eliash_sigma})
and~(\ref{eq.g_dyson}), we fully include
the effect of the level shift function, 
which comes from the antisymmetric part of the energy dependent DOS~\cite{carbotte_tc}.
The level shift function and the self-consistency might be crucial for
superconducting property
when the DOS has a strong energy dependence around the Fermi level. 
Through the change of the normal Green's function, the pairing
interaction in Eq.\,(\ref{eq.eliash_delta}) could be modified~\cite{pickett_prb_82,carbotte_cjp_83}. 
We will discuss this point in
Sec.\,\ref{sec.eliash}. 
%%% to arita pickett たちの 仕事自身から self-

%More practically, the momentum
%average of $\tilde{V}^{\mathrm{ph}}$ is performed with the delta function weight for
%the orbitals near the Fermi level.
%By using the weight function, the averaged electron-phonon matrix
%element is written as
More practically, 
the momentum average of $\tilde{V}^{\mathrm{ph}}$ and the averaged electron-phonon 
matrix elements are calculated as 
\begin{eqnarray}
 \label{eq.g_average}
  |g^{jl}_{\lambda} ({\bm q})|^2
  = \frac{\sum_{\bm{p}} 
  |g^{j\bm{p}+\bm{q},l\bm{p}}_{\lambda} ({\bm q})|^2 
  \delta(\xi_{j\bm{p}+\bm{q}}) \delta(\xi_{l\bm{p}})}{
  \sum_{\bm{p}} \delta(\xi_{j\bm{p}+\bm{q}}) \delta(\xi_{l\bm{p}})}.
\end{eqnarray}
%and if the averaged $|g|^2$ is smaller than the threshold value 
%(as in the case of bands far from the Fermi level ),
%$g^{jl}_{\lambda} ({\bm q})$ is evaluated by simple averaging.
When the $j$- and $l$-th band are far away from the Fermi level and 
$g^{jl}_{\lambda} ({\bm q})$ evaluated by \eqref{eq.g_average} is smaller than a threshold value, 
the averaged matrix element is approximately calculated as 
\begin{align}
 |g^{jl}_{\lambda} ({\bm q})|^2
 = \frac{1}{N}
 \sum_{\bm{p}} 
 |g^{j\bm{p}+\bm{q},l\bm{p}}_{\lambda} ({\bm q})|^2.
\end{align}
%It is known that the electron-phonon coupling sometimes is not accurately
%evaluated without sophisticated
%interpolation~\cite{mauri_prl,akashi_prb,giustino_prb_07,kawamura_opt_tetra}.
%This is generally due to subtle dependence of the joint density of
%states $\sum_{\bm{p}} \delta(\xi_{j\bm{p}+\bm{q}})
%\delta(\xi_{l\bm{p}})$ on the integration method. 
%In this work, we expect that averaging with the weight mitigates the
%possible error in the pairing strength without any special interpolation.

%Another strong point of this treatment is about the convergence for the
%Matsubara frequency summation. 
As mentioned earlier, the advantage of solving Eqs.\,\eqref{eq.eliash_sigma} 
and~\eqref{eq.eliash_delta} is the inclusion of the energy dependence of DOS.  
Another advantage 
is that we can explicitly treat the retardation effect.
In the calculation with the constant DOS approximation, 
it is in principle impossible to achieve the convergence with respect to the number of
%% to arita ここ the いるんですか?
Matsubara frequencies.
Since %several numerical calculations show that
the gap function is almost constant in the high frequency region~\cite{rainer_tc}, 
%increase of the number of the Matsubara frequency in Eq.\,(\ref{eq.gap_iso}) leads
%to the logarithmic divergence, if $\mu^*$ is fixed. (Of course $\mu^*$ is
%cutoff dependent object in true meaning.) 
the second term in the r.h.s.~of Eq.\,(\ref{eq.gap_iso}) always diverges 
logarithmically for fixed $\mu$ and sufficiently large number 
of Matsubara frequencies. 
To avoid this logarithmic divergence, $\mu$ must be zero, 
which is totally unphysical.
Therefore, one cannot reach a converged solution for non-empirically
calculated $\mu$ without introducing an adjustable electron energy
cutoff~\cite{mu_star}. %in realistic materials. 
%On the other hand, energy dependent treatment can mitigate such problem
%since in Eq.\,(\ref{eq.eliash_delta}), the paring interaction is
%proportional to $G_{j\bm{p}}(i\omega_n) G_{j-\bm{p}}(-i\omega_n)$ and it
%behaves like $1/\omega_n^2$ at high frequency regions. 
Such a problem is mitigated in Eq.\,(\ref{eq.eliash_delta}), since 
$G_{j\bm{p}}(i\omega_n) G_{j-\bm{p}}(-i\omega_n)$ behaves as $1/\omega_n^2$
in the high frequency limit.
%High-\tc of
%the sulfur hydries also helps the convergence naturally since covered
%frequency ranges are enlarged by raising temperature. Therefore, one can
%calculate \tc from first principles based on this formulation.
However, the numerical cost to achieve the convergence with respect to 
the Matsubara frequency is formidable, 
especially when \tc is low.
This is because we need an extremely large number of Matsubara frequencies 
to cover the high frequency region.
Fortunately, the \tc's of sulfur hydrides are $30$-$200$\,K,
which enables the Matsubara frequency grid to span a wide range of energy 
with a small number of grid points. 
We can therefore carry out the \tc calculation non-empirically--without introducing 
the adjustable parameters--with a feacible numerical cost.  
%calculate \tc nonempirically, i.e., 
%without introducing the adjustable parameter $\mu^*$.

In the calculation with the constant DOS approximation, 
one can also obtain the converged \tc with fixed $\mu$ by introducing
another cutoff frequency $\omega_{\rm el}$ for effective energy range of
the Coulomb interaction, instead of $\omega_c$. 
With taking the isotropic limit and linearization, 
the momentum summation of the r.h.s.~in Eq.\,\eqref{eq.eliash2} can be
calculated analytically
\begin{align}
 &\ \ \ \ \ \frac{1}{N}\sum_{\bm{p}' l} G_{l \bm{p}'}(i\omega_{n'}) G_{l
 -\bm{p}'}
 (-i\omega_{n'}) \nonumber  \\
& = N(0) \int_{-\omega_{\rm el}}^{\omega_{\rm el}} \diff \xi
 \frac{1}{ Z(i\omega_{n'})^2 \omega_{n'}^2 + \xi^2}  \nonumber \\
& =  \frac{2N(0)}{Z(i\omega_{n'}) \omega_{n'}} \arctan \left(
 \frac{\omega_{\rm el}}{ Z(i\omega_{n'}) \omega_{n'} } \right).
\end{align}
Thus, Eq.\,\eqref{eq.gap_iso} becomes 
\begin{align}
 \label{eq.gap_iso_cutoff}
 \phi(i\omega_n)
  = \frac{1}{Z(i\omega_n)} \frac{\pi}{\beta}
  \sum_{n'} \frac{\phi(i\omega_{n'})}{|\omega_{n'}|}
  \biggl\{& \lambda(i\omega_n - i\omega_{n'})  \nonumber \\
 &  \ \ \ - \mu \eta_{n'} (\omega_{\rm el})  \biggr\},
\end{align} 
where $\eta_{n} (\omega_{\rm el})$ is the cutoff function defined by 
\begin{align}
 \eta_{n} (\omega_{\rm el}) = \frac{2}{\pi} 
 \arctan \left( \frac{\omega_{\rm el}}{ Z(i\omega_{n}) |\omega_{n}| } 
 \right).
\end{align}
Here, although the effective energy range is considerably different, 
both $\lambda(i\omega_n)$ and $\eta_{n} (\omega_{\rm el})$ decay
as a function of $\omega_n$.
By combining Eqs.\,\eqref{eq.Z_iso} and \eqref{eq.gap_iso_cutoff}, 
one can calculate \tc with the fully non-empirically evaluated $\mu$ 
if a large number of Matsubara frequencies is taken.
%and the empirical parameter $\omega_{\rm el}$ is introduced. 
Hereafter we call this treatment constant DOS ME theory.

%%%%%%%%%%%%%%%%%%%%%%%%%%%%%%
\subsection{Allen-Heine-Cardona theory}

The Allen-Heine-Cardona (AHC) theory is a perturbative approach to calculate
ZPR from first-principles~\cite{ahc_76,ahc_81,giustino_prl_10,gonze_11}. 
If the Hamiltonian $H$ is perturbed by
ion displacement $u$ from its equilibrium position, it causes a shift of the
electron energy. At the level of the second order perturbation, such a shift is given by
\begin{align}
 \delta \epsilon_{j\bm{p}} 
  = \frac{1}{2N} &\sum_{\kappa \kappa', {\bm q} \lambda, \mu\nu} \sum_{ll'}
  \sqrt{ \frac{\hbar^2}{ M_\kappa M_{\kappa'} \omega_{{\bm q}\lambda}^2 }} 
   \nabla_{l\kappa\mu}\nabla_{l'\kappa'\nu} \epsilon_{j\bm{p}} \nonumber \\
  &\times e_{\kappa}^{\mu*} ({\bm q} \lambda)   e_{\kappa'}^{\nu} ({\bm q}
  \lambda) 
  e^{i{\bm q} \cdot ({\bm R}_{l'} - {\bm R}_{l})}
  \biggl\{ \langle n_{{\bm q} \lambda} \rangle + \frac{1}{2} \biggr\}
  \nonumber \\
  = \frac{1}{2N} &\sum_{\kappa \kappa', {\bm q} \lambda, \mu\nu}
  \sqrt{ \frac{\hbar^2}{ M_\kappa M_{\kappa'} \omega_{{\bm q}\lambda}^2
  }} 
  e_{\kappa}^{\mu*} ({\bm q} \lambda)e_{\kappa'}^{\nu} ({\bm q}
  \lambda) \nonumber \\
  &\times \frac{\partial^2 \epsilon_{j \bm{p}}}
   {\partial u^*_{\mu\kappa}(\bm{q}) \partial u_{\nu \kappa'}(\bm{q}) } 
  \biggl\{ \langle n_{{\bm q} \lambda} \rangle + \frac{1}{2} \biggr\}
\end{align}
where $e^{\mu}_{\kappa}(\bm{q}\lambda)$
is the phonon polarization vector with momentum
$\bm{q}$ and mode $\lambda$ defined through
Eq.\,\eqref{eq.dynamical_eigen}, 
$N$ is the number of ${\bm q}$-points, 
$M_{\kappa}$ is the mass of the $\kappa$-th ion, 
$\bm{R}_l$ is the position of the $l$-th unit cell, 
$\langle n_{{\bm q} \lambda} \rangle$ is the Bose-Einstein distribution
function, 
%$u$ is the displacement of ions, 
and 
$\nabla_{l\kappa\mu}\nabla_{l'\kappa'\nu} \epsilon_{j \bm{p}}$ is the second
order derivative of the Kohn-Sham energy~\cite{gonze_prb_14} defined by
\begin{align}
 \label{eq.2nd_derivative}
  \nabla_{l\kappa\mu}\nabla_{l'\kappa'\nu} \epsilon_{j \bm{p}}
 &\ \  = \langle \psi_{j\bm{p}} | \nabla_{l\kappa\mu}\nabla_{l'\kappa'\nu}
 H | \psi_{j \bm{p}}
  \rangle \nonumber \\
 & + \left\{ \langle \nabla_{l'\kappa'\nu}\psi_{j \bm{p}} |
  \nabla_{l\kappa\mu} H | \psi_{j \bm{p}}
  \rangle +  {\rm c.c.} \right\} %\langle \psi_{\bm{p}j} | \nabla_{\kappa\mu} H | \nabla_{\kappa'\nu} \psi_{\bm{p}j} \rangle \}
\end{align}
with the Kohn-Sham orbital $\psi_{j\bm{p}}$ and the Kohn-Sham energy
$\epsilon_{j\bm{p}}$.
$\nabla_{l\kappa\mu}$ represents the derivative with respect to the the $\kappa$-th
ion position in the $l$-th unit cell for the $\mu$-th direction.  
The shift of the band energy coming from the first order and second order derivatives are called
the Fan term~\cite{fan} and Debye-Waller (DW) term~\cite{dw}, respectively. 
Here, the first order modulation of the Hamiltonian can be obtained by
DFPT~\cite{dfpt}. 
To evaluate the first order derivative of the wave function, one can
utilize the Sternheimer approach~\cite{gonze_11,gonze_cms_14,gonze_prb_14} by separating the unoccupied manifold
from the occupied space by
\begin{eqnarray}
  |\nabla_{l\kappa\mu}\psi_i \rangle = 
  - \sum_{j; {\rm occ}}   
  \frac{\langle \psi_j | \nabla_{l\kappa\mu} H | \psi_i \rangle
  }{\epsilon_i - \epsilon_j} | \psi_j \rangle \nonumber \\
  + P_{\rm unocc} | \nabla_{l\kappa\mu}\psi_i \rangle, 
\end{eqnarray}
where the first term of the r.h.s.~can be calculated by summing only over
the occupied states, 
and the second term can be evaluated by standard DFPT~\cite{dfpt}. 
Here $P_{\rm unocc}$ is projection
to the unoccupied manifold. For the details about DFPT, see Appendix.

Compared with the first order derivative of the Hamiltonian, %it is quite
%difficult to calculate the second order one since it needs much
%more computational costs. 
the second order derivative requires much more computational costs.
The first order derivative can be treated as
%%% to arita monotonic を monochromatic に
monochromatic perturbation~\cite{dfpt}, which means that for the calculation
at momentum $\bm{q}$, it does not need information about other momenta
$\bm{q}'\neq\bm{q}$. On the other hand, calculation for the second
order derivative is not monochromatic and needs additional loop for momentum. 
%To our knowledge, there is no work calculating
%$\langle \psi_{j \bm{p}} | \nabla_{l\kappa\mu}\nabla_{l'\kappa'\nu} H |
%\psi_{j \bm{p}} \rangle$
%for infinite solids directly. 
%In order to avoid the direct calculation of the second order derivative, 
Instead of calculating 
$\langle \psi_{j \bm{p}} | \nabla_{l\kappa\mu}\nabla_{l'\kappa'\nu} H |
\psi_{j \bm{p}} \rangle$
directly, 
one usually employs the acoustic sum rule and the rigid-ion
approximation~\cite{ahc_76}.
The acoustic sum rule represents the fact that the uniform displacements of
the ions have no effect on the periodic system. It gives the
constraint which connects the second order derivative with the first
order through the following equation~\cite{gonze_prb_14}:
\begin{eqnarray}
 \label{eq.sum_rule}
 \sum_{\kappa'}  \frac{\partial^2 \epsilon_{j \bm{p}}}
 {\partial u^*_{\mu\kappa}(\bm{0}) \partial u_{\nu \kappa'}(\bm{0}) }
 = 0.
\end{eqnarray}

%In addition to the acoustic sum rule, 
On top of that, one can use the rigid-ion approximation. 
%one can achieve the replacement of
%the second order derivative of the Hamiltonian 
Namely, 
one can replace the second order derivative of the Hamiltonian
with the first order derivative
if the Hamiltonian is assumed to have the following form:
\begin{eqnarray}
 \label{eq.rigid-ion_H}
  H_{\rm{rigid\mathchar`-ion}} = 
  K + \sum_{l\kappa} V_{l\kappa}(\bm{r} - \bm{R}_{l}- \bm{r}_{\kappa})
\end{eqnarray}
with the electron kinetic energy $K$ and the potential energy
$V_{l\kappa}$ caused by the $\kappa$-th ion in the $l$-th unit cell. 
Here $\bm{r}_{\kappa}$ and $\bm{R}_{l}$ represent the position of 
the $\kappa$-th ion and the $l$-th unit cell, respectively.
%This is the rigid-ion approximation. 
It is
apparent that $\nabla_{l\kappa\mu}\nabla_{l'\kappa'\nu} H$ is equal to
zero if $\kappa \neq \kappa'$ as well as $l \neq l'$. This property
leads to the following formula:
\begin{align}
 \label{eq.rigid-ion}
 &\left\langle \psi_{j \bm{p}} \left|  \left. \left. \frac{\partial^2 H}
 {\partial u^*_{\mu\kappa}(\bm{q}) \partial u_{\nu \kappa'}(\bm{q})} \right. \right.\right|
  \psi_{j \bm{p}} \right\rangle \nonumber \\
&\ \ \ \  =  \left\langle \psi_{j \bm{p}} \left| \left. \left. \frac{\partial^2 H}
 {\partial u^*_{\mu\kappa}(\bm{0}) \partial u_{\nu \kappa'}(\bm{0}) }
 \right. \right. \right|
  \psi_{j \bm{p}} \right\rangle \delta_{\kappa\kappa'}.
\end{align}
%Combining Eq.\,(\ref{eq.rigid-ion}) with Eq.\,(\ref{eq.sum_rule}),
Combining Eqs.\,(\ref{eq.2nd_derivative}),~(\ref{eq.sum_rule}) and~(\ref{eq.rigid-ion}), 
one can evaluate the shift of the electron band energy by
\begin{eqnarray}
 \delta \epsilon_{j \bm{p}} = \frac{1}{N} \sum_{\bm{q} \lambda} 
  \frac{\partial \epsilon_{j \bm{p}}}{\partial n_{\bm{q}\lambda}} 
  \biggl\{ \langle n_{{\bm q} \lambda} \rangle + \frac{1}{2} \biggr\},
\end{eqnarray} 
where $\partial \epsilon_{j \bm{p}} /\partial n_{\bm{q}\lambda}$ is
divided into two contributions
\begin{eqnarray}
 \frac{\partial \epsilon_{j \bm{p}}}{\partial n_{\bm{q}\lambda}} 
 = \frac{\partial \epsilon^{(\rm Fan)}_{j \bm{p}}}{\partial
 n_{\bm{q}\lambda}}
 + \frac{\partial \epsilon^{(\rm DW)}_{j \bm{p}}}{\partial n_{\bm{q}\lambda}},
\end{eqnarray}
and each term is written as
\begin{align}
% \frac{\partial \epsilon^{(\rm Fan)}_{j \bm{p}}}{\partial
% n_{\bm{q}\lambda}} 
%& =& \frac{\hbar}{2\omega_{\bm{q}\lambda}} 
% \sum_{\kappa\kappa', \mu\nu} \sqrt{\frac{1}{M_\kappa M_\kappa'}} 
%  e_{\kappa}^{\mu*} ({\bm q} \lambda)e_{\kappa'}^{\nu} ({\bm q}
%  \lambda) \nonumber \\
% &\times& \frac{1}{2} \biggl\{ \bigl\{ \langle \frac{\partial \psi_{j \bm{p}}}{\partial
%  u_{\kappa\mu} (\bm{q})} |
%  \frac{\partial H}{\partial
%  u_{\kappa' \nu} (\bm{q})}  | \psi_{j \bm{p}}
%  \rangle  \nonumber \\ 
% &+& (\kappa\mu) \leftrightarrow (\kappa'\nu) \bigr\} + {\rm c.c.}
 % \biggr\}. 
 \label{eq.fan}
 \frac{\partial \epsilon^{(\rm Fan)}_{j \bm{p}}}{\partial
 n_{\bm{q}\lambda}} 
& = \frac{\hbar}{2\omega_{\bm{q}\lambda}} 
 \sum_{\kappa\kappa', \mu\nu} \sqrt{\frac{1}{M_\kappa M_{\kappa'}}} 
  e_{\kappa}^{\mu*} ({\bm q} \lambda)e_{\kappa'}^{\nu} ({\bm q}
  \lambda) \nonumber \\
 &\ \ \times  \biggl\{ \left\langle \frac{\partial \psi_{j \bm{p}}}{\partial
  u_{\kappa\mu} (\bm{q})} \left|
   \left. \left.  \frac{\partial H}{\partial
  u_{\kappa' \nu} (\bm{q})} \right. \right.  \right| \psi_{j \bm{p}}
  \right\rangle  
 + {\rm c.c.} \biggr\} 
\end{align}
\begin{align}
% \frac{\partial \epsilon^{(\rm DW)}_{j \bm{p}}}{\partial
% n_{\bm{q}\lambda}} 
% & =& -\frac{\hbar}{4\omega_{\bm{q}\lambda}} 
% \sum_{\kappa\kappa', \mu\nu} 
% \frac{1}{2} \biggl\{ \bigl\{ \langle \frac{\partial \psi_{j \bm{p}}}{\partial
%  u_{\kappa\mu} (\bm{0})} |
%  \frac{\partial H}{\partial
%  u_{\kappa' \nu} (\bm{0})}  | \psi_{j \bm{p}}
%  \rangle  \nonumber \\ 
% &+& (\kappa\mu) \leftrightarrow (\kappa'\nu) \bigr\} + {\rm c.c.} \biggr\}. 
% \biggl\{ \frac{e_{\kappa}^{\mu*} ({\bm q} \lambda)e_{\kappa}^{\nu} ({\bm q}
%  \lambda)}{M_\kappa}
% + \frac{e_{\kappa'}^{\mu*} ({\bm q} \lambda)e_{\kappa'}^{\nu} ({\bm q}
%  \lambda)}{M_\kappa'}  \biggr\}
%   \nonumber \\
% \label{eq.dw}
% \frac{\partial \epsilon^{(\rm DW)}_{j \bm{p}}}{\partial
% n_{\bm{q}\lambda}} 
%  = &-\frac{\hbar}{4\omega_{\bm{q}\lambda}} 
% \sum_{\kappa\kappa', \mu\nu} 
% \biggl\{ \left\langle \frac{\partial \psi_{j \bm{p}}}{\partial
%  u_{\kappa\mu} (\bm{0})} \left|
%  \left. \left. \frac{\partial H}{\partial
%  u_{\kappa' \nu} (\bm{0})} \right. \right. \right| \psi_{j \bm{p}}
%  \right\rangle  
% + {\rm c.c.} \biggr\} \nonumber \\
% &\times \biggl\{ \frac{e_{\kappa}^{\mu*} ({\bm q} \lambda)e_{\kappa}^{\nu} ({\bm q}
%  \lambda)}{M_\kappa}
% + \frac{e_{\kappa'}^{\mu*} ({\bm q} \lambda)e_{\kappa'}^{\nu} ({\bm q}
%  \lambda)}{M_\kappa'}  \biggr\}.
 \label{eq.dw}
 \frac{\partial \epsilon^{(\rm DW)}_{j \bm{p}}}{\partial
 n_{\bm{q}\lambda}} 
 & =  -\frac{\hbar}{4\omega_{\bm{q}\lambda}} 
 \sum_{\kappa\kappa', \mu\nu} 
 \biggl\{ \frac{e_{\kappa}^{\mu*} ({\bm q} \lambda)e_{\kappa}^{\nu} ({\bm q}
  \lambda)}{M_\kappa}
 + \frac{e_{\kappa'}^{\mu*} ({\bm q} \lambda)e_{\kappa'}^{\nu} ({\bm q}
  \lambda)}{M_{\kappa'}}  \biggr\} \nonumber \\
 &\ \ \ \times\biggl\{ \left\langle \frac{\partial \psi_{j \bm{p}}}{\partial
  u_{\kappa\mu} (\bm{0})} \left|
  \left. \left. \frac{\partial H}{\partial
  u_{\kappa' \nu} (\bm{0})} \right. \right. \right| \psi_{j \bm{p}}
  \right\rangle  
 + {\rm c.c.} \biggr\}.
\end{align}

Here let us consider the validity of the rigid-ion approximation. 
In the Kohn-Sham system, the Hamiltonian does not have the form of
Eq.\,(\ref{eq.rigid-ion_H}) due to the Hartree and exchange-correlation
potential.
These potential terms depend on the electron density, and the electron
density response to the displacement of one ion is affected by that of
other ions. Therefore, the potential term cannot be expressed as the
sum of the potentials of the individual ions. 
In spite of this fact, one can still expect that the rigid-ion approximation
works well in three-dimensional materials as a consequence of the electronic
screening~\cite{dw_rigidion}. %in the case of the DW term. 
%The screening makes the effective length of the displacement of
%the different ions short and the existence of the such complex
%movement is visible only near the edge of the Brillouin zone (BZ). 
%%% to arita the existence of the such complex movement を such short range effect に 変更 まだ こっちのほうが わかりやすい?
The screening makes the range where the displacement of the different 
ions affects small and such short range effects is relevant only near the
edge of the Brillouin zone (BZ).
Since the volume of the edge regions becomes small in higher dimensional systems, 
one could safely apply the rigid-ion approximation to
three-dimensional materials. 
For the DW contribution, this statement is confirmed in the case of
diamond by comparing the result of the AHC theory with that of the frozen
phonon approach~\cite{gonze_prl_14}.

In this study, we employ the AHC theory implemented in {\sc abinit}~\cite{abinit}
package for the evaluation of the ZPR. 
%{\sc abinit} is an {\it ab initio}
%calculation package based on the density functional theory with plane wave
%pseudopotential method. 
In the {\sc abinit} calculation, we use the same
pseudopotential used in the electronic and phononic structure
calculations (see Sec.\,\ref{sec.elph}).

%%%%%%%%%%%%%%%%%%%%%%%%%%%%%%%%%%%%%%%%%%%%%%%%%%%%%%
\subsection{Self-consistent phonon theory}

%Anharmonicity causes several effects on the phonon properties.
%Important effects for superconductivity are the change of the phonon
%dispersion, the distortion of the phonon polarization vector, and the
%existence of the finite lifetime of the phonons. Although any of these
%effects are related to \tc, we only consider the change of the phonon
%dispersion. In this study we focus on pressure regions far from structural
%transition point. Since anharmonicity would be less important in such
%regions, this treatment could be good as simple estimation.
%
%
%In order to consider the effect of the anharmonicity, 
%we employ the self-consistent phonon theory (SCPT)~\cite{tadano_jpsj,tadano_prb}. 

%Anharmonicity causes several effects on the phonon properties.
%Important effects for superconductivity are the change of the phonon
%dispersion, the distortion of the phonon polarization vector, and the
%existence of the finite lifetime of the phonons. Although any of these
%effects is related to \tc, we only consider the change of the phonon
%dispersion. In this study, we focus on pressure regions far from structural
%transition point. Since anharmonicity would be less important in such
%regions, this treatment could be good as simple estimation.

In this paper, we study how the anharmonicity changes the phonon dispersion 
and affects the self-energy of electrons, and consequently the 
superconducting \tc. 
Several \textit{ab initio} approaches have been recently proposed for including
anharmonic effects of phonons beyond quasiharmonic level~\cite{souvatzis_scaild,
hellman_tdep,errea_sscha,needs_anharmonic,tadano_prb}. Here, we employ a
deterministic method based on the self-consistent phonon (SCPH)
theory~\cite{werthamer_scph,tadano_prb}. In our approach, the first-order effect
of the frequency renormalization  due to the quartic anharmonicity is treated
nonperturbatively by solving the following SCPH equations:
\begin{align}
&\det{\{\omega^{2}-\bm{U}_{\bm{q}}\}} =0, \label{eq.scph_det} \\
&U_{\bm{q}\lambda\lambda'}=\omega_{\bm{q}\lambda}^{2} \delta_{\lambda\lambda'}
+(2\omega_{\bm{q}\lambda})^{\frac{1}{2}}(2\omega_{\bm{q}\lambda'})^{\frac{1}{2}}
\Pi_{\bm{q}\lambda\lambda'}. \label{eq.scph}
\end{align}
Here the matrix $\bm{\Pi}_{\bm{q}}$ is the lowest-order phonon self-energy 
associated with the quartic terms defined as
\begin{equation}
\Pi_{\bm{q}\lambda\lambda'} = \sum_{\bm{q}_{1},\lambda_{1}}
\frac{\hbar\Phi(\bm{q}\lambda;-\bm{q}\lambda';\bm{q}_{1}\lambda_{1};-\bm{q}_{1}\lambda_{1})}
{8\sqrt{\omega_{\bm{q}\lambda}\omega_{\bm{q}\lambda'}}\omega_{\bm{q}_{1}\lambda_{1}}}
\left[ 2\langle n_{\bm{q}_{1}\lambda_{1}} \rangle + 1 \right].
\label{eq.scph_selfenergy}
\end{equation}
The tensor $\Phi$ in the numerator represents the strength of the phonon-phonon coupling 
and can be calculated from the fourth-order interatomic force constants (IFCs) in real space as follows:
\begin{align}
&\Phi(\bm{q}\lambda;-\bm{q}\lambda';\bm{q}_{1}\lambda_{1};-\bm{q}_{1}\lambda_{1}) \nonumber \\
&= \frac{1}{N}\sum_{\{\kappa,\mu,l\}}
\frac{e_{\kappa_{1}}^{\mu_{1}}(\bm{q}\lambda)e_{\kappa_{2}}^{\mu_{2}*}(\bm{q}\lambda')e_{\kappa_{3}}^{\mu_{3}}(\bm{q}_{1}\lambda_{1})e_{\kappa_{4}}^{\mu_{4}*}(\bm{q}_{1}\lambda_{1})}
{\sqrt{M_{\kappa_{1}}M_{\kappa_{2}}M_{\kappa_{3}}M_{\kappa_{4}}}}\nonumber \\
&\hspace{10mm} \times \Phi_{\mu_{1}\mu_{2}\mu_{3}\mu_{4}}(0\kappa_{1};l_{2}\kappa_{2};l_{3}\kappa_{3};l_{4}\kappa_{4}) \nonumber \\
&\hspace{10mm} \times e^{-i[\bm{q}\cdot\bm{R}_{l_{2}}-\bm{q}_{1}\cdot(\bm{R}_{l_{3}}-\bm{R}_{l_{4}})]}.
\label{eq.scph_quartic}
\end{align}
By diagonalizing the Hermitian matrix $\bm{U}_{\bm{q}}$, one obtains phonon frequencies and 
corresponding eigenvectors modulated by fourth-order anharmonicity. 
Through this change of phonon frequencies and eigenvectors, the phonon self-energy $\bm{\Pi}_{\bm{q}}$ 
is also updated. Therefore,
Eqs.~(\ref{eq.scph_det})--(\ref{eq.scph_quartic}) need to be solved
iteratively 
until a convergence is achieved with respect to anharmonic phonon frequencies.
In this study, we neglect the mode off-diagonal elements of the phonon self-energy, 
i.e. $\Pi_{\bm{q}\lambda\lambda'}\approx \Pi_{\bm{q}\lambda\lambda}\delta_{\lambda,\lambda'}$,
so that the phonon polarization vectors are not altered by anharmonic effects.
The SCPH solution also includes the effect of zero-point motion [Eq.\,(\ref{eq.scph_selfenergy})],
which is crucial for understanding anharmonic effects in sulfur hydrides under pressure~\cite{mauri_prl}.

To conduct the SCPH calculation, we need to calculate the fourth-order IFCs. 
For that purpose, we employ the real-space supercell approach, and anharmonic force constants are extracted from 
displacement-force training data sets prepared by DFT calculations. 
To reduce the number of independent IFCs and make the computation feasible, 
we make full use of space group symmetries and constraints due to the translational invariance~\cite{tadano_jpsj}.
Moreover, we employ the compressed sensing lattice dynamics method~\cite{zhou_lasso} for 
reliable and efficient estimation of force constants. An efficient implementation and 
more technical details of the present SCPH calculation can be found in Ref.\,\cite{tadano_prb}. 

\section{Electronic and phononic structure} \label{sec.elph}

In previous works, several stable structures under high pressures are
determined~\cite{ma_jpc,duan_SciRep,mazin_prb,duan_prb_15,gross_arxiv,akashi_prb,mauri_prl,ma_arxiv_15,mauri_arxiv_15,akashi_hxs}. 
Here we focus on the difference between \hhs and
\hhhs. 
In order to avoid the difficulty coming from 
structure instability near the transition point, we choose the pressures
far from the critical pressure, which is around $180$\,GPa in \hhhs 
and around $160$\,GPa in \hhs.
For pressures close to the critical point, the electron-phonon coupling is
strongly enhanced due to the structure instability, especially in the case
of the second order phase transition in
\hhhs~\cite{duan_SciRep,akashi_prb, mauri_prl,mauri_arxiv_15}, 
which considerably raises \tc.
However, such enhancement of \tc might be an artifact of the
harmonic approximation which cannot be justified in the vicinity of 
the phase transition since there are huge ion oscillations toward other
stable structure and one cannot assume the amplitude of these
oscillations to be small.
In Ref.\,\cite{mauri_prl}, it is reported that 
anharmonicity strongly suppresses the electron-phonon coupling especially
near the transition point for \hhhs.
Therefore, we hereafter choose $250$\,GPa for \hhhs and $140$\,GPa for \hhs.

In order to study the superconducting property, first one should
obtain the electronic and phononic structure of the target material
precisely. 
%For such purpose, 
%density functional theory is a strong tool to calculate the
%electronic structure from first principles.
For phonon frequencies and
electron-phonon matrix elements, we utilize the framework of
DFPT~\cite{dfpt}  
as implemented in {\sc quantum espresso}~\cite{qe}. 
%which is an {\it
%ab initio} calculation package based on the density functional theory
%with plane wave pseudopotential method.
Density functional calculations are performed within the generalized gradient
approximation using the Perdew-Burke-Ernzerhof parameterization~\cite{pbe96}.
Atomic configurations and lattice constants are optimized by minimizing
enthalpy under fixed pressures. 
%Details on the stable structures are listed in
%Ref.\,\cite{akashi_prb}.

\subsection{Electronic structure}

\begin{figure}[htbp]
\vspace{0cm}
 \centering
  \includegraphics[width=0.35\textwidth]{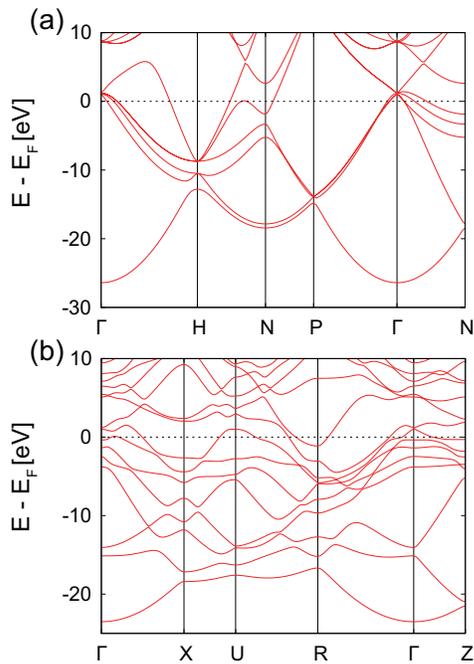}
 \caption{Band structures of (a) $Im\bar{3}m$-\hhhs at $250$\,GPa
 and (b) $P\bar{1}$-\hhs at $140$\,GP along
 several high symmetry lines. Energy is measured from the Fermi level.}
  \label{fig_band}
\end{figure}

In Fig.\,\ref{fig_band}, we show the band dispersion of \hhhs at
$250$\,GPa and \hhs at $140$\,GPa.
Under these pressures, $Im\bar{3}m$ and $P\bar{1}$ are the energetically most stable
structures for \hhhs and \hhs, respectively. 
The electron charge densities are obtained with $16\times16\times16$ BZ mesh for
$Im\bar{3}m$ and $12\times12\times8$ BZ mesh for $P\bar{1}$. 
The cutoff for the plane-wave energy is set to $100$\,Ry ($80$\,Ry) for
$Im\bar{3}m$ ($P\bar{1}$). 
We use the pseudopotential implemented based on the Troullier-Martins 
scheme~\cite{tm_pseudo}. 
In both Fig.\,\ref{fig_band} (a) and (b), the electronic bands 
far below the Fermi level have free electron like parabolic dispersion.
However, near the Fermi level, there is a notable difference 
between $Im\bar{3}m$ and $P\bar{1}$ structures. 
%Since $P\bar{1}$ is a molecular crystal, the electron dispersion is relatively flat compared with the band in $Im\bar{3}m$ structure.

\begin{figure}[htbp]
\vspace{0cm}
 \centering
  \includegraphics[width=0.45\textwidth]{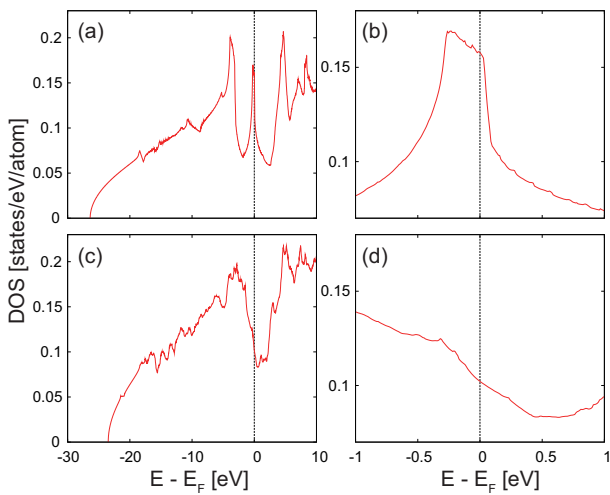}
 \caption{Densities of states of (a) $Im\bar{3}m$-\hhhs and (c)
 $P\bar{1}$-\hhs. 
 (b) and (d) show the
 enlarged views of (a) and (c) around the Fermi level, respectively. 
 The unit of the vertical axis is states per an atom. 
% We use the same charge distribution for the band calculation
% in Fig.\,\ref{fig_band}. 
 In (b), there is a
 sharp peak around the Fermi level and the peak width is comparable with the phonon
 energy scale, while (d) does not show any characteristic structures within
 the energy range of 1\,eV.}
 \label{fig_dos}
\end{figure}

To clarify the difference of the electronic structures 
between $Im\bar{3}m$-\hhhs and $P\bar{1}$-\hhs,
%the DOS of these two structures are shown in Fig.\,\ref{fig_dos}.
%Here these DOS are calculated with the tetrahedron
%method~\cite{tetrahedron}.
the DOS calculated by the tetrahedron method~\cite{tetrahedron} for 
these two structures are shown in Fig.\,\ref{fig_dos}. 
%As seen in the band dispersions, there are free electron like structures
%below the Fermi level, over the range from $-30$\,eV to $-5$\,eV in both phases.
Near the Fermi level, there is a dip in the DOS for the $P\bar{1}$ structure.
In Fig.\,\ref{fig_dos} (b) and (d), the enlarged views of the DOS are shown. 
In this energy scale, the DOS of the $P\bar{1}$ structure is almost flat. 
On the other hand, there is a strong energy dependence of the DOS as
a consequence of the vHs around the Fermi level for the 
$Im\bar{3}m$ structure (Fig.\,\ref{fig_dos} (b)). This narrow peak is a
characteristic
common feature of the DOS of the \hhhs phases since one can also observe it 
in the $R3m$ structure~\cite{duan_SciRep}, another stable structure 
stabilized under pressures around $180$\,GPa~\cite{duan_SciRep,duan_prb_15,akashi_prb,ma_arxiv_15}.

The existence of the vHs is a good news for high-\tc
superconductivity since the large DOS at the Fermi level enhances 
the electron-phonon coupling~\cite{pickett_arxiv}.
However, the vHs makes it difficult to treat the superconductivity
theoretically since the constant DOS approximation is not justified. 
This is the same situation as in the A15
compounds~\cite{carbotte_ssc_78,pickett_prb_82,pickett_prl_82,carbotte_cjp_83,carbotte_prb_83}. 
The constant DOS approximation overestimates
the number of relevant states for superconductivity  around the Fermi
level and consequently \tc~\cite{gross_arxiv,pickett_arxiv,pickett_prb_82}.

%%%%%%%%%%%%%%%%%%%%%
\subsection{Phononic structure}

\begin{figure}[htbp]
 \vspace{0cm}
 \centering
  \includegraphics[width=0.45\textwidth]{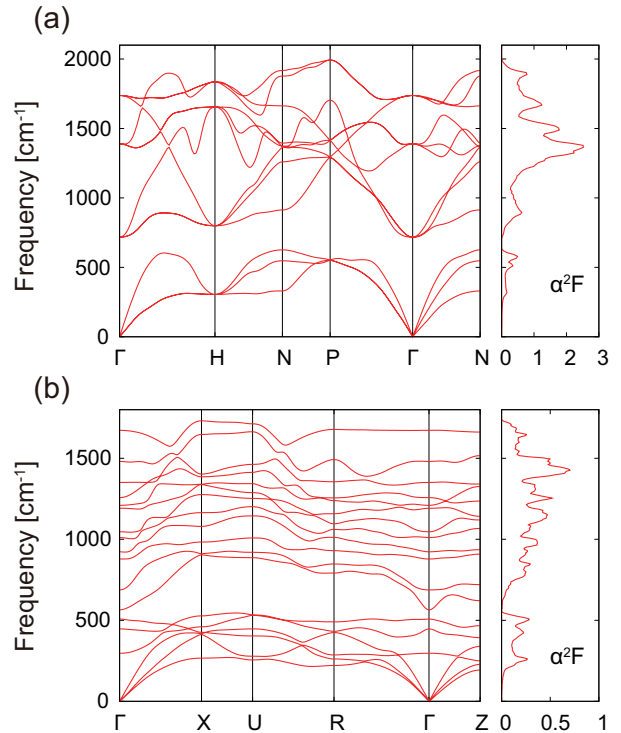}
 \caption{Phonon dispersions and Eliashberg functions $\alpha^2 F
 (\omega)$ in (a) $Im\bar{3}m$-\hhhs and (b) $P\bar{1}$-\hhs.}
  \label{fig_disp}
\end{figure}

Figure \ref{fig_disp} shows the phonon dispersion relations for the two
structures. 
For the linear response calculation, we use $10\times10\times10$ ${\bm q}$-mesh
for $Im\bar{3}m$-\hhhs and $12\times12\times8$ ${\bm q}$-mesh
for $P\bar{1}$-\hhs, respectively.
It is important to notice that the typical phonon energy scale 
is extraordinarily high.
In both Fig.\,\ref{fig_disp}\,(a) and\,(b), the frequencies of the hardest
modes are above $1500$\,cm$^{-1}$.
This value is larger than phonon energies in simple metals by a factor of
ten.
The existence of these hard phonons is consistent with the prediction of
high-\tc superconductivity in hydrogen-rich
compounds~\cite{ashcroft_Hrich}.

To give more quantitative discussion about the energy scale of the
phonons, we calculate $\omega_{\rm ln}$ defined by
\begin{eqnarray}
 \label{eq.omega_ln}
 \omega_{\rm ln} = \exp \biggl\{ 
  \frac{2}{\lambda} \int_{0}^{\infty} \diff \omega \frac{\alpha^2
  F(\omega)}{\omega} \ln \omega \biggr\},
\end{eqnarray}
where $\lambda$ is the electron-phonon coupling constant defined by $\lambda(0)$
in Eq.\,(\ref{eq.elph}). The value of $\omega_{\rm ln}$ is $987$\,K
($686$\,cm$^{-1}$) for the $P\bar{1}$ structure
and $1521$\,K ($1057$\,cm$^{-1}$) for the $Im\bar{3}m$ structure. 
These values are consistent with previous works
~\cite{ma_jpc,duan_SciRep,gross_arxiv,akashi_prb,mauri_prl}.
%Since the hardness of phonons help to raise
%\tc, \hhhs composition is favorable as a candidate of high-\tc superconductor.
Here it should be noticed that the scale of $\omega_{\rm ln}$ in \hhhs
is comparable with the peak width of the DOS (Fig.\,\ref{fig_dos}\,(b)).
It clearly indicates that one should seriously consider the energy dependence of DOS to
study the superconducting properties. %since $\omega_{\rm ln}$ shows the
%typical phonon energy scale.

%In addition to the high frequency phonons, the electron-phonon coupling
%also tells us that \hhhs should have higher \tc than \hhs.
%$\lambda$ takes the values of $0.86$ in $P\bar{1}$-\hhs, whereas it reaches
%$1.83$ in $Im\bar{3}m$-\hhhs at $250$\,GPa. 
%These values, which are calculated using Eq.\,(\ref{eq.elph}) with averaged
%%%% to koretsune coupling は 不可算名詞 ？
%electron-phonon matrix elements given by Eq.\,(\ref{eq.g_average}), 
%are also consistent with previous studies~\cite{akashi_prb,mauri_prl}.
%%Such values are large enough to treat both phases as strong coupling
%%superconductors (In general sense, $0.86$ is still large enough).
%%The BCS theory suggests that Large electron-phonon coupling raises \tc thorough the
%%enhancement of pairing interaction.
%Therefore, %these two phases are expected to be superconductors, and
%the $Im\bar{3}m$ structure is expected to have higher \tc. %as a consequence of both the hardness
%%of the phonons and the greatness of the electron-phonon coupling.
%%This discussion also works qualitatively for other stable structures
%For other stable structures, such predictions of \tc also work qualitatively
%($R3m$ for \hhhs~\cite{duan_SciRep} and $cmca$ for \hhs~\cite{ma_jpc}).

In addition to the high frequency phonons, the electron-phonon coupling
also tells us that \hhhs should have higher \tc than \hhs.
$\lambda$ takes the values of $0.86$ in $P\bar{1}$-\hhs, whereas it reaches
$1.83$ in $Im\bar{3}m$-\hhhs at $250$\,GPa 
with the first-order Hermite-Gaussian
approximation~\cite{m-p_smear} for the delta functions with the smearing 
width of $0.010$\,Ry.
These values, which are calculated using Eq.\,(\ref{eq.elph}) with averaged
electron-phonon matrix elements given by Eq.\,(\ref{eq.g_average}), 
are also consistent with previous studies~\cite{akashi_prb,mauri_prl}
($\lambda = 1.96$ in Ref.\,\cite{mauri_prl} with the Wannier
interpolation for the electron-phonon matrix elements~\cite{giustino_wan},
and $\lambda = 1.97$ in Ref.\,\cite{akashi_prb} with the optimized
tetrahedron method~\cite{kawamura_opt_tetra} for the electron delta function 
in Eq.\,\eqref{eq.eliash_function} in $Im\bar{3}m$-\hhhs).
Therefore, 
the $Im\bar{3}m$ structure is expected to have higher \tc.
For other stable structures, such predictions of \tc also work qualitatively
($R3m$ for \hhhs~\cite{duan_SciRep} and $Cmca$ for \hhs~\cite{ma_jpc}).

%As mentioned in Sec.\,\ref{sec.method}, $\lambda$ sometimes has errors 
%with coarse mesh sampling~\cite{mauri_prl,akashi_prb}. 
%We also calculate $\lambda$ with fully momentum dependent
%electron-phonon matrix elements. In the \hhhs-$Im\bar{3}m$ structure at $250$\,GPa, we
%obtain $\lambda = 1.73$ with the first-order Hermite-Gaussian
%approximation~\cite{m-p_smear} for the delta functions with the smearing 
%width of $0.030$\,Ry.
%This value is smaller than that of previous studies
%($\lambda = 1.96$ in Ref.\,\cite{mauri_prl} with the Wannier
%interpolation for the electron-phonon matrix elements~\cite{giustino_wan},
%and $\lambda = 1.97$ in Ref.\,\cite{akashi_prb} with the optimized
%tetrahedron method~\cite{kawamura_opt_tetra} for the electron delta function 
%in Eq.\,\eqref{eq.eliash_function}).
%By using Eq.\,(\ref{eq.g_average}), $\lambda$ is increased up to $1.83$ and it is
%confirmed that the
%discrepancy between ours and previous works is mitigated. 
%%On the other hand, $\omega_{\rm ln}$ only has weak dependence to
%%calculation meshs or interpolation methods.
%On the other hand, $\omega_{\rm ln}$ does not depend much on the calculation meshes 
%or interpolation methods.

Here we discuss why there is a huge difference in the electron-phonon coupling
between \hhhs and \hhs. 
One reason comes from the DOS. 
Since $\lambda$ is roughly proportional to the DOS at the Fermi level, 
the electron-phonon coupling is enhanced by the large DOS (see Eq.\,(\ref{eq.eliash_function})).
From Fig.\,\ref{fig_dos}\,(b) and\,(d), the DOS takes larger value in $Im\bar{3}m$-\hhhs
than in $P\bar{1}$-\hhs. 
Through the larger number of the available states around the Fermi level,
$\lambda$ in the $Im\bar{3}m$ structure should become larger. %(One should take care of the unit of DOS).
%Another point is the bonding character of hydrogen atoms.
%One can see that the dispersion of the hard phonons (above $1000$\,cm$^{-1}$) is
%nearly flat in $P\bar{1}$ structure (Fig.\,\ref{fig_disp}\,(b)). 
%Since high frequency phonon is almost completely contributed by hydrogen
%atoms, it means that $P\bar{1}$ structure has the molecular type bonding
%characters for
%hydrogen atoms and the coupling between hydrogen vibration modes and
%conduction electron is relatively weak. 
%This character can be seen in Fig.\,5 of Ref.\,\cite{ma_jpc}. 
%It shows that the half of $\lambda$ comes from the low-lying sulfur vibration (six
%modes) and
%others comes from the hydrogen oscillation.
%One the other hands, high frequency phonons in $Im\bar{3}m$ structure is
%strongly dispersive due to the covalent character of the hydrogen-sulfur
%bond~\cite{pickett_prb,mazin_prb}.
%Consequently, the large part of $\lambda$ (about $70$\,\%) is contributed by high energy
%hydrogen vibration modes (See Fig.\,5 in Ref.\,\onlinecite{duan_SciRep}).
Another point is the coupling strength between the electrons and 
the hydrogen vibration.
It is shown that a half of $\lambda$ comes from the 
low-lying sulfur vibrations (six modes below $500$\,cm$^{-1}$) and
the rest comes from the hydrogen oscillation in \hhs~\cite{ma_jpc}.
%On the other hand, the high frequency hydrogen oscillations mainly contribute to
%the electron-phonon coupling in the \hhhs phases~\cite{duan_SciRep}
On the other hand, $\lambda$ in \hhhs originate mainly from
the high frequency hydrogen oscillations. 
($70$\,\% of $\lambda$ is contributed from 
the hydrogen bond-stretching phonons~\cite{mazin_prb}).
Therefore, between \hhhs and \hhs, there is a clear
difference in the coupling of the electrons and the hydrogen oscillating
phonons. 
It is pointed out that the larger $\lambda$ in \hhhs is ascribed to 
the strong covalency of the hydrogen-sulfur bonding 
in Ref.\,\cite{mazin_prb}.

\section{Effect of strong el-ph coupling on the van Hove singularity and
 superconductivity} \label{sec.eliash}

Using the electronic and phononic structure calculation in
Sec.\,\ref{sec.elph},
we perform the calculation of \tc for $Im\bar{3}m$-\hhhs and $P\bar{1}$-\hhs
based on the ME theory.
As mentioned in Sec.\,\ref{sec.intro}, 
%it should be checked the
%availability of the additional approximations
%since it might ignore important characters of the systems. 
%In addition to the justification of the calculation methods, the ZPR should
%be taken into account due to the lightness of the hydrogen atoms. 
for sulfur hydrides, we have to go beyond the constant DOS approximation
employed in the previous
calculations~\cite{ma_jpc,duan_SciRep,li_arxiv,mauri_prl,ma_arxiv_15,pietronero_arxiv_15_2,mauri_arxiv_15}.
In this section, we will show that the energy dependence of the DOS is
indeed crucial to describe the retardation effect properly. 
%While the Green's function and self-energy are not calculated
%self-consistently in the constant DOS approximation, 
%we also discuss how the self-consistency is important for the
%quantitative estimate of \tc. 
We also discuss how the self-consistency in the Green's function and
self-energy is important for the quantitative estimation of \tc, 
which is not considered in the previous ME approaches with the constant
DOS approximation.
As is recently suggested by Ref.\,\cite{arxiv_bianconi}, 
in sulfur hydrides, 
the effect of ZPR on the spectral function can be significant. 
In this section, we also examine how ZPR affects superconductivity.
Let us discuss these points one by one in the following subsections.

\subsection{Energy dependence of DOS}
In this subsection, we examine the importance of the energy dependence
of the DOS for the accurate description of the retardation effect. 
With the strong energy dependence of the DOS at the Fermi level, 
it is expected
%that this approximation does not work for the \hhhs phase, although
%it's not so bad assumption for the \hhs phases.
that the constant DOS approximation is more problematic in \hhhs
than in \hhs.
%In order to confirm this speculation, 
%In order to see how the breakdown of this approximation affects both
%phases, 
In order to see the effect of the energy dependence of DOS on the both
compounds, 
we compare \tc calculated by
Eqs.\,(\ref{eq.Z_iso}) and~(\ref{eq.gap_iso_cutoff}) with that by
Eqs.\,(\ref{eq.eliash_sigma}) and~(\ref{eq.eliash_delta}).
%Here one should take care of the self-consistency for the normal Green's
%function.
%Although the ME theory with energy dependent DOS can utilize the SC normal
%Green's function for the gap equation, 
%it is impossible to solve Eq.\,(\ref{eq.gap_iso_cutoff}) with the self-consistent
%renormalization function.
%%the constant DOS ME theory is solved
%%with the non-SC renormalization function inevitably.
%To extract the difference resulting from the constant DOS approximation, 
%we employ the non-SC normal Green's function in both methods.
%The effect of the self-consistency will be discussed in the next subsection.
While the self-consistent dressed Green's function should be used in
Eq.\,\eqref{eq.eliash_delta}, 
here we employ the one-shot Green's function to focus on the effect of
the energy dependence of the DOS on the retardation effect.
It should be noticed that the self-consistency of the Green's function
is not taken into account in the constant DOS approximation with
Eqs.\,\eqref{eq.Z_iso} and~\eqref{eq.gap_iso_cutoff}.

%In the ME theory, it is important to consider how to include the Coulomb
%interaction.
In the calculation based on the ME theory, 
the numerical cost to treat the Coulomb interaction is generally very expensive.
This is because the Coulomb interaction is effective in the whole range of the 
band width, 
and thus it requires a large number of Matsubara frequencies. % to cover a wide energy range.
Therefore, one needs special care for the convergence with respect to
the cutoff for the Matsubara frequency sum.

\begin{figure}[htbp]
\vspace{0cm}
 \centering
  \includegraphics[width=0.32\textwidth]{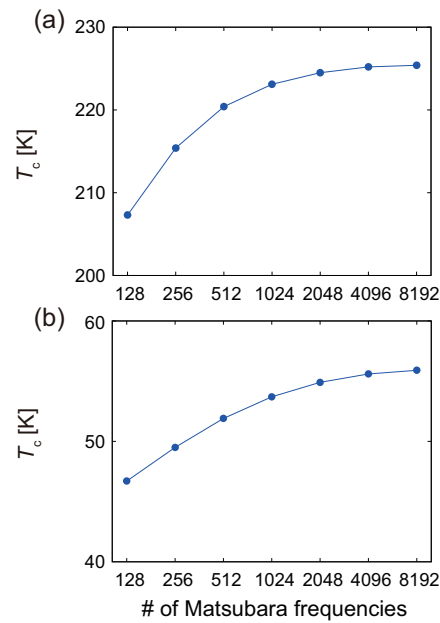}
\vspace{0cm}
 \caption{\tc against the number of Matsubara frequencies for
 (a) $Im\bar{3}m$-\hhhs at $250$\,GPa and (b)
 $P\bar{1}$-\hhs at $140$\,GPa within the constant DOS
 approximation. 
 \tc is calculated for fixed $\mu$ values of $\mu = 0.32$ in (a) and
 $\mu = 0.16$ in (b) evaluated with the RPA.
 The value of $\omega_{\rm el}$ is fixed at $20$\,eV for the both
 structures.}
% \tc is calculated within the constant DOS approximation for
% $\mu=0.32$ in (a) and $\mu=0.16$ in (b).
% The values of $\mu$ is evaluated with the RPA.}
 \label{fig_omega_constdos}
\end{figure}
\begin{figure}[htbp]
\vspace{0cm}
\centering
  \includegraphics[width=0.32\textwidth]{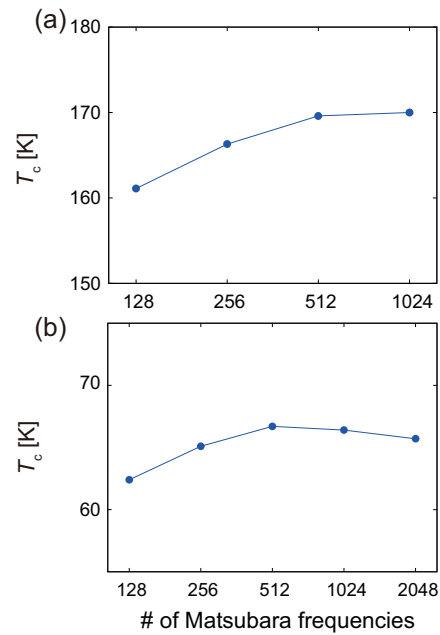}
\vspace{0cm}
 \caption{Number of Matsubara frequencies dependence of \tc for
 (a) $Im\bar{3}m$-\hhhs at $250$\,GPa and (b)
 $P\bar{1}$-\hhs at $140$\,GPa 
  based on the ME theory with energy dependent DOS.
 The screened Coulomb interaction is evaluated with the RPA.}
 \label{fig_omega_nscf_bandU}
\end{figure}

%By the RPA with density functional theory, one can obtain the screened Coulomb
%interaction from first principles.
%For the calculation with energy dependent DOS, we can simply apply the obtained
%Coulomb interaction for Eq.\,(\ref{eq.eliash_delta}) directly (although
%we ignore the frequency dependence of the Coulomb interaction).
%On the other hand, it is necessary to evaluate the averaged and
%renormalized Coulomb potential $\mu^*$ for the constant DOS ME theory. 
%However, the average of the RPA screened Coulomb interaction is directly
%connected with not $\mu^*$ but the bare Coulomb potential $\mu$.
%%Here words of 'bare' and 'renormalize' denotes the 
%Here, it is natural to expect that $\mu$ is the pesudo Coulomb potential with the limit
%$\omega_c \rightarrow \infty$ in Eq.\,(\ref{eq.gap_iso_cutoff}).
Let us examine this problem in the calculation within and beyond the
constant DOS approximation. 
As is discussed in Sec.\,\ref{sec.method}, 
we employ the RPA for the screened Coulomb interaction in 
Eq.\,\eqref{eq.rpa_coulomb}. 
Following the argument by Migdal and Eliashberg~\cite{schrieffer}, 
we neglect the frequency dependence of the RPA screened Coulomb
interaction: $\tilde{V}^{\mathrm c} (i\omega_n) = \tilde{V}^{\mathrm c}(0)$. 
In the constant DOS calculation with Eqs.\,\eqref{eq.Z_iso}
and~\eqref{eq.gap_iso_cutoff}, 
we introduce the averaged Coulomb potential $\mu$
(Eq.\,\eqref{eq.mu_coulomb}) 
and the adjustable cutoff $\omega_{\rm el}$. 
One can calculate \tc with non-empirically evaluated $\mu$ by
utilizing $\omega_{\rm el}$ and the cutoff function. 
Figure~\ref{fig_omega_constdos} shows 
the summation cutoff dependence
%the number of Matsubara frequencies dependence 
of \tc with fixed $\mu$. 
The value of $\omega_{\rm el}$ is set as $20$\,eV for the both structures. 
If the cutoff frequency $\omega_{\rm el}$ is fixed, 
we can obtain converged results with tractable numbers of Matsubara
frequencies. 
It is also confirmed that the same results can be obtained by using
Eq.\,\eqref{eq.gap_iso} as the gap equation, 
where $\mu^*$ is calculated by employing Eq.\,\eqref{eq.mustar_coulomb}
with empirically selected $\omega_{\rm el}$ and $\omega_{\rm c}$. 
Here, it should be noticed that this calculation is not fully {\it ab
initio} because of the introduction of the cutoff for the effective
Coulomb energy range. 
Although \tc is converged with fixed $\omega_{\rm el}$, 
there still remains $\omega_{\rm el}$ dependence of \tc~\cite{w_el_comment}. 
Now let us move on to the calculation beyond the constant DOS
approximation to see how it solves the problem of the Matsubara
frequency sum. 
In order to cover wide energy range around the Fermi level, 
we include $7$ ($12$) bands from the band bottom in
$Im\bar{3}m$-\hhhs ($P\bar{1}$-\hhs). 
As mentioned in Sec.\,\ref{sec.method}, Eq.\,(\ref{eq.eliash_delta}) has
natural convergent factor $1/\omega_n^2$ in the frequency sum. 
Figure~\ref{fig_omega_nscf_bandU} shows the Matsubara frequency dependence
of \tc in the calculation with energy dependent DOS. 
Here the normal Green's function is calculated by the one-shot treatment.
%In contrast to the case of the constant DOS approximation, a converged
%\tc
A converged \tc is obtained with 512 Matsubara frequencies, slightly
smaller number than that in the constant DOS calculation. 
This treatment has great advantage compared with the constant DOS ME
theory  
since we can evaluate \tc by
the fully non-empirical calculation based on the ME theory. 
%%It should be noticed that conventional ME theory is not even fully
%%nun-empirical with adjustable parameter $\mu^*$ and frequency cutoff $\omega_c$.
%Hereafter, in order to compare results obtained by two methods, 
%we set the summation cutoff for the constant DOS ME theory as
%$1024$ Matsubara frequency for the \hhhs phases and $2048$ Matsubara
%frequency for the \hhs phases.

Finally, let us compare the results of the calculations with constant
DOS and energy dependent DOS. 
We set the summation cutoff as $8192$ frequencies in the constant DOS
calculation for both structures, 
and $1024$ and $2048$ frequencies with energy dependent DOS for 
\hhhs and \hhs, respectively. 
Figure~\ref{fig_omega_constdos}\,(a) shows that \tc is $225$\,K with
the constant DOS approximation, 
and it decreases to $168$\,K without the constant DOS
approximation for $Im\bar{3}m$-\hhhs  
(Fig.\,\ref{fig_omega_nscf_bandU}\,(a)).
\tc falls by 
$57$\,K ($34$\,\%) by considering the energy dependence of DOS. 
It clearly indicates that the constant DOS approximation breaks down for
$Im\bar{3}m$-\hhhs~\cite{gross_arxiv}.
On the other hand, 
as seen in Fig.\,\ref{fig_omega_constdos}\,(b)
and~\ref{fig_omega_nscf_bandU}\,(b), \tc increases only by $10$\,K ($15$\,\%)
in $P\bar{1}$-\hhs 
(from $56$\,K with constant DOS to $66$\,K). 
By comparing the results in \hhhs and \hhs, we can conclude 
that the existence of the
narrow peak at the Fermi level 
must be treated carefully for the accurate estimate of \tc. 
%strongly affects \tc, and the energy
%dependence of DOS should play a crucial role for the superconducting
%state in \hhhs. 

%%%%%%%%%%%%%%%%%%%%%
\subsection{Self-consistency for the normal Green's function}

\begin{figure}[htbp]
\vspace{0cm}
 \centering
  \includegraphics[width=0.35\textwidth]{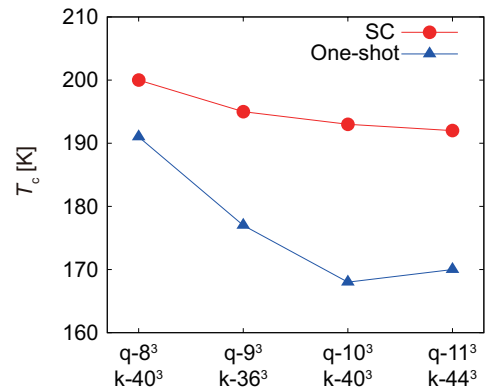}
 \caption{Numerical convergence of \tc in $Im\bar{3}m$-\hhhs at
 $250$\,GPa. SC (red line) and One-shot (blue line) denote the
 self-consistent and one-shot
 calculations for the normal Green's function. 
 The number of Matsubara frequencies is fixed at $1024$.
 ${\bf q}$ and ${\bf k}$ represent the Brillouin zone sampling employed for the phonon dynamical
 matrix and the gap equation, respectively.}
  \label{fig_mesh_sh3}
\end{figure}

In this subsection, we discuss the importance of the self-consistency for the
normal Green's function.
We perform the calculation of \tc with both the one-shot and SC
treatment for the normal Green's function. 
Figure~\ref{fig_mesh_sh3} shows the mesh dependence of
\tc in $Im\bar{3}m$-\hhhs.
In both cases, convergence is already achieved with
$10\times10\times10$ mesh for the phonon calculation.

\begin{figure}[htbp]
\vspace{0cm}
 \centering
  \includegraphics[width=0.40\textwidth]{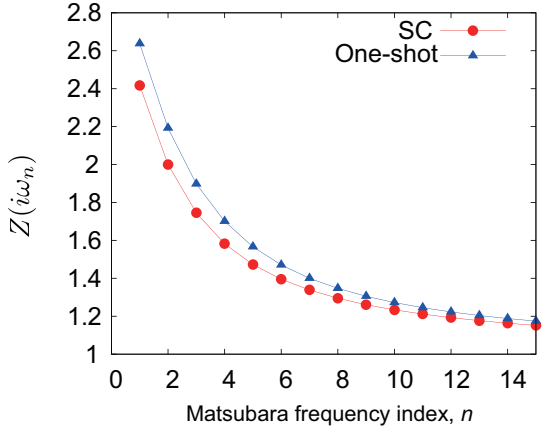}
 \caption{Renormalization function $Z(i\omega_n)$ as a function of the Matsubara
 frequency for $Im\bar{3}m$-\hhhs at $250$\,GPa.
 Here the Matsubara frequency is defined by $\omega_n = (2 n - 1)\pi/\beta$.
 %The band near the Fermi level is selected (fifth band counted from
 %lowest band).
 We choose the band crossing the Fermi level (the fifth band from
 the band bottom).
 SC (red line) and One-shot (blue line) denote the self-consistent and
 one-shot calculation for the normal Green's function, respectively.
 Plotted $Z$ is calculated at $185$\,K in the SC calculation and $166$\,K
 in the one-shot calculation.
 Temperature dependence of $Z$ is weak and irrelevant.
 %and there is only subtle change in these plots for higher temperatures.
 }
 \label{fig_renorm}
\end{figure}

The results are listed in the second and third row in Table\,\ref{tab.tc}.
There is a clear difference of \tc in the case of $Im\bar{3}m$-\hhhs.
\tc is enhanced by $25$\,K through the self-consistency of
the normal Green's function. One can see that such enhancement of \tc
is accompanied by the reduction of the renormalization function. In
Fig.\,\ref{fig_renorm}, the momentum averaged renormalization
function defined as 
\begin{align}
 Z(i\omega_n) = 1 - \frac{\mathrm{Im} \Sigma(i\omega_n)}{\omega_n}
\end{align}
is plotted as a function of the Matsubara frequency. 
In both cases, $Z$ is suppressed and approached to one in the limit of
large $\omega_n$. 
By contrast, near the minimum Matsubara frequency, $Z$ in the one-shot treatment 
takes larger value than that in the SC calculation. 
%The self-consistency might mitigate the development of $Z$.
There is a feedback effect in the self-consistent loop which mitigates
the development of $Z$.
The ratio of $Z(i \pi/\beta )$ by the SC calculation to the one-shot calculation is $0.915$ just below
the transition temperature.
The suppression of the renormalization function leads to the enhancement
of \tc through the pairing interaction since $Z$ denotes the mass
enhancement of electrons.
%Therefore, the increase of \tc occurs by the reduction of $Z$ at low
%frequency region.

%We can expect that the suppression of the renormalization function by
%the self-consistency is also related with the vHs,
On the other hand, \tc does not show such huge variation in the case of
$P\bar{1}$-\hhs. 
Here we also calculate the ratio of $Z(i \pi/\beta)$ for $P\bar{1}$-\hhs and find 
%that it is nearly one (the value is $0.987$).
that it is $0.987$, which is much closer to unity than that of \hhhs in
the $Im\bar{3}m$ structure. 
%Combined with the fact that \tc does not differ between SC and
%one-shot calculation in the \hhs phase, 
This result indicates that the large suppression of the renormalization
function by the self-consistency in \hhhs is related with the existence of
the vHs. 
%Therefore, it is confirmed that the self-consistency should
It reveals that the self-consistency is 
another important factor for accurate calculation of \tc when the DOS
has strong energy dependence.

%\subsection{Zero-point renormalization for superconductivity}
\subsection{Effect of zero-point motion on superconductivity}

%Before discussing the effect of ZPR, we should think of how to include
%the energy shift by zero-point motion for the calculation of \tc.
%Although each AHC and ME theory works individually, there is a double
%counting problem if these are directly combined. 
%That is because the diagram of the Fan term 
%in the AHC theory (Eq.\,(\ref{eq.fan})) is
%equivalent to the diagram considered in the ME theory. 
%This problem is safely avoided by omitting the Fan contribution in the
%calculation of ZPR.
%Therefore, after that we treat only the DW contribution (Eq.\,(\ref{eq.dw}))
%as ZPR.

\begin{figure}[htbp]
\vspace{0cm}
 \centering
  \includegraphics[width=0.40\textwidth]{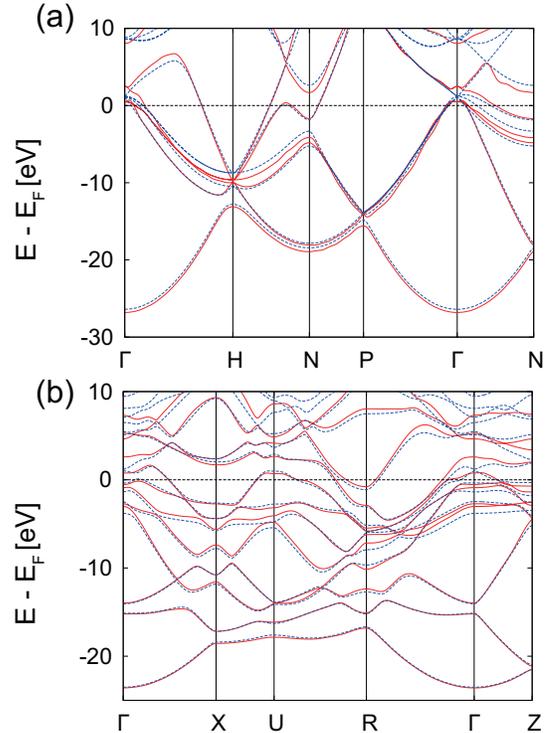}
 \caption{Band dispersions with (red solid line) and without (blue broken
 line) ZPR for (a) $Im\bar{3}m$-\hhhs at $250$\,GPa
 and (b) $P\bar{1}$-\hhs at $140$\,GPa. 
 %Zero-point renormalization is
 %evaluated based on Allen-Heine-Cardona theory implemented in ABINIT
 %packages.
 Band structures with ZPR are plotted by the Wannier interpolation.}  
 \label{fig_band_zpr}
\end{figure}

\begin{figure}[htbp]
 \vspace{0.2cm}
 \centering
 \includegraphics[width=0.30\textwidth]{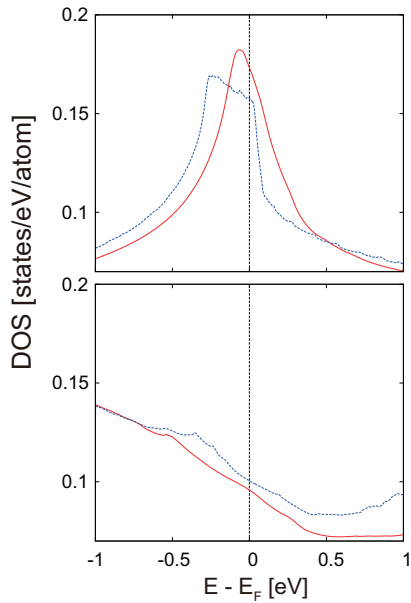}
 \vspace{0cm}
 \caption{Densities of states with (red solid line) and 
 without (blue broken line) ZPR for
 (a) $Im\bar{3}m$-\hhhs at $250$\,GPa and (b)
 $P\bar{1}$-\hhs at $140$\,GPa near the Fermi level.}
 \label{fig_dos_zpr}
\end{figure}

Figure~\ref{fig_band_zpr} shows how the ZPR changes the band dispersions for
\hhhs and \hhs.
Since the Fan term in the AHC theory (Eq.\,\eqref{eq.fan}) is already
considered in the ME theory (Eq.\,\eqref{eq.eliash1}), hereafter we only
take account of the contribution of the DW term (Eq.\,\eqref{eq.dw}) as
ZPR. 
%The harmonic phonon frequency is used in the evaluation of the DW term. 
For the calculation of the ZPR, $18\times18\times18$ and
$14\times14\times8$ BZ sampling are employed for 
$Im\bar{3}m$-\hhhs and $P\bar{1}$-\hhs, respectively. 
Temperature dependence of the band energy shift is small and only leads to slight changes
in the calculation of \tc. 
%As mentioned above, we only include the DW term in this figure.
In Fig.\,\ref{fig_band_zpr}, %one might think that there are only
			       %small energy shifts
there is apparently only small energy shifts 
by zero-point motion so that the ZPR in sulfur hydrides is not important.
However, it is not necessarily the case 
%since energy shifts of order of hundred meV
%can have huge effects on superconducting states. 
since the energy shifts amount to a few hundred meV, 
which can generally have significant effects on superconductivity. 
%In order to clarify the shifts of the electronic structure within the
%important energy ranges, let us take a closer look at the
%DOS. 
%In Fig.\,\ref{fig_dos_zpr}, the density of states
%around the Fermi level is shown.
Let us now take a closer look at the DOS in the energy range relevant to
superconductivity shown in Fig.\,\ref{fig_dos_zpr}.
It clearly indicates that the DOS drastically changes due to the ZPR
within 
%such a small energy range.
%It is this energy scale which affects the superconducting property since
%the phonon energy scale is of order of $100$\,meV in both $Im\bar{3}m$ and $P\bar{1}$
%structure.
the energy scale of phonons.
It is interesting to note that 
there is an enhancement (suppression) of the DOS around the Fermi level
for \hhhs (\hhs).

With the ZPR, we %repeatedly perform the calculation of \tc.
calculate \tc.
We treat the electron energy dispersion with the ZPR as an input for 
the ME calculation.
The results are shown in the forth row in Table\,\ref{tab.tc}.
In both $Im\bar{3}m$-\hhhs and $P\bar{1}$-\hhs, 
there is a shift of \tc %by about
%$10$\,K, 
of the order of ten kelvin, 
although the direction is opposite. 
In the $Im\bar{3}m$ structure, \tc gets raised by the ZPR, 
while it decreases in the $P\bar{1}$ structure.
%Such shifts have the same directions with the changes of the DOS shown in
%Fig.\,\ref{fig_dos_zpr}.
Such shifts are naively consistent with the changes of the DOS shown in
Fig.\,\ref{fig_dos_zpr}. 
In fact, we obtain $\lambda=2.06$ with ZPR and $1.83$ without ZPR for
$Im\bar{3}m$-\hhhs, and $\lambda=0.73$ with ZPR and
$0.86$ without ZPR for $P\bar{1}$-\hhs. 

%The change of DOS at the Fermi level affects the electron-phonon
%coupling through the increase/decrease of the scattering path for
%superconducting pairing.
%Therefore, one should take a look at the variation of $\lambda$ by ZPR.
%By including the ZPR, we obtain $\lambda=2.06$ for the \hhhs-$Im\bar{3}m$ structure at
%$250$\,GPa. This value is larger than $1.83$ without the ZPR.
%Although due to the vHs, it is dangerous to relate the electron-phonon coupling
%directly with \tc in the \hhhs phases, this enhancement of
%$\lambda$ denotes the development of the pairing interaction.
%On the other hand, $\lambda$ decreases from $0.86$ to $0.73$ by
%the suppression of the DOS in the case of the \hhs-$P\bar{1}$ structure.
%This fact suggests that \tc is affected by the ZPR through the change of
%available states near the Fermi level.
%%We couldn't ignore the ZPR in the case of sulfur hydrides since the
%%magnitude of \tc shift is of order of $10$\,K in both structures.

Here it should be noted that there is a difference in the ratio of
\tc shifts to the original values between \hhhs and \hhs. 
\tc is raised by $9$\,K ($5$\,\%) in $Im\bar{3}m$-\hhhs, 
while there is a decrease of \tc by $19$\,K ($30$\,\%) in the
$P\bar{1}$-\hhs. 
This results indicate that, in \hhhs,  the enhanced pairing
interaction is smeared out within the energy scales of phonons and 
the effects of the attractive interaction are weaker than that expected
by the value of $\lambda$.  
On the other hand, in \hhs, 
since the DOS is reduced within the energy scales relevant to
superconductivity, 
the effects of the reduction of the pairing interaction remain large
even if the smearing effects are taken into account.
%!TEX root = paper2.tex

\section{Anharmonicity}
\label{sec.anharm}

\begin{figure}[htbp]
\vspace{0cm}
 \centering
 \includegraphics[width=0.40\textwidth]{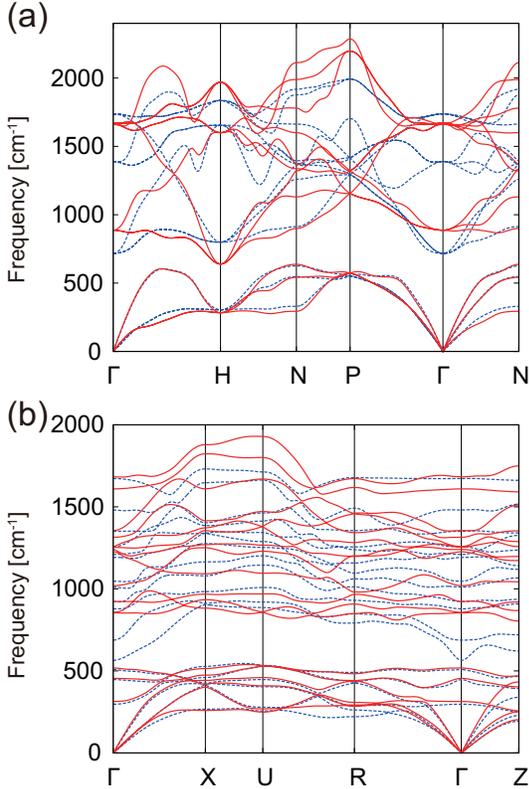}
 \caption{Phonon dispersion relation for (a) $Im\bar{3}m$-\hhhs
 at $250$\,GPa and (b) $P\bar{1}$-\hhs at $140$\,GPa. 
 The red solid line shows the
 dispersion considering the anharmonic effect within the SCPH theory, and the blue broken line shows the
 result within the harmonic approximation.}
 %Anharmonic dispersion is calculated using the self-consistent phonon
 %theory with $10\times10\times10$ mesh for the dynamical matrix and
 %$5\times5\times5$ mesh for  the self-consistent phonon theory equation. }
 \label{fig_disp_ah}
\vspace{0cm}
\end{figure}

We also examine the anharmonic effect on the calculation of \tc. 
In the previous study based on the constant DOS approximation for
\hhhs~\cite{mauri_prl}, 
it has been shown that the anharmonicity makes 
%the phonon frequency higher, and 
the electron-phonon coupling weaker. 
Consequently, \tc dramatically decreases especially for the pressure
near which the system undergoes the structural transition. 
Here we study the anharmonic effect in \hhs and \hhhs considering the
energy dependence of the DOS. 
%Figure~\ref{fig_disp_ah} shows the anharmonic phonon dispersion of the
%\hhhs-$Im\bar{3}m$
%and \hhs-$P\bar{1}$ structures calculated by using the SCPH theory.
Figure~\ref{fig_disp_ah} shows the phonon dispersion of
$Im\bar{3}m$-\hhhs under $250$\,GPa and that of $P\bar{1}$-\hhs under
$140$\,GPa. 
Here we compare the results obtained by the SCPH theory (red solid lines)
and by the harmonic approximation (blue dotted lines).

The SCPH calculations were conducted as follows:
First, we performed first-principles molecular dynamics (FPMD) simulations at 300 K using
$3\times3\times3$ and $3\times3\times2$ supercells for $Im\bar{3}m$-\hhhs and $P\bar{1}$-\hhs, respectively, and extracted physically relevant atomic configurations in every 50 MD steps ($\sim$ 24.2 fs).
For the sampled snapshots, we then added random displacements to each atom
to reduce the cross-correlation inherent in the FPMD trajectories~\cite{zhou_lasso}. 
For \hhhs (\hhs), we prepared 40 (250) displacement patterns and calculated forces for each configuration using {\sc quantum espresso}. 
Next, using the displacement-force training data, we estimated anharmonic IFCs by the least absolute shrinkage and selection operator (LASSO) method. 
Here, anharmonic IFCs up to the sixth order were included in the anharmonic lattice model \cite{tadano_jpsj} to improve the prediction accuracy. 
The total number of independent parameters was as large as 9000 (28000) for \hhhs (\hhs), from which a sparse solution having $\sim$ 3700 ($\sim$ 16000) non-zero parameters was obtained by LASSO with a regularization parameter determined by cross-validation. 
The accuracy of the estimated IFCs was checked by applying them to independent test configurations, 
where the interatomic forces predicted by the model showed good agreements with DFT values for both \hhhs and \hhs.
Finally, we solved the SCPH equation [Eqs.~(\ref{eq.scph_det})-(\ref{eq.scph_quartic})] at 0 K using the quartic IFCs estimated by LASSO and the harmonic dynamical matrices calculated by DFPT. 
The SCPH equation was solved using $5\times5\times5$ $\bm{q}$-mesh for \hhhs and $3\times3\times2$ $\bm{q}$-mesh for \hhs, 
and the same mesh densities were employed for the $\bm{q}_{1}$ point in Eq.~(\ref{eq.scph_selfenergy}). Doubling the $\bm{q}_{1}$-mesh points along each direction did not change the results for both systems, indicating the finite size effect is not significant at the selected pressures.
The anharmonic correction to the dynamical matrix $\Delta \bm{D}(\bm{q})=\bm{D}^{\mathrm{SCPH}}(\bm{q})-\bm{D}^{\mathrm{DFPT}}(\bm{q})$ was transformed into the real-space force constants $\Delta \Phi_{\mu\nu}(0\kappa;l'\kappa')$,
from which anharmonic phonon frequencies at denser $\bm{q}$ points were obtained by interpolation.

We see that the anharmonicity changes the phonon dispersion, 
especially for phonons whose frequencies are higher than $\sim
500$\,cm$^{-1}$~\cite{mauri_prl}.  
As a result, the electron-phonon coupling is weakened from $2.06$ to
$1.86$ ($1.83$ to $1.64$) for the \hhhs,
and from $0.73$ to $0.64$ ($0.86$ to $0.75$) for the \hhs when the ZPR is
considered (neglected)~\cite{zpr_comment}.  
By contrast, $\omega_{\rm ln}$ stays nearly unchanged by the
anharmonicity (from $1521$\,K ($987$\,K) with the harmonic approximation, to
$1515$\,K ($1034$\,K) with the anharmonicity for $Im\bar{3}m$-\hhhs
($P\bar{1}$-\hhs)).

The modification of the phonon frequencies gets \tc to be lowered. 
It is confirmed by the calculation of \tc based on the self-consistent ME theory
with the zero-point renormalization.
As listed in Table\,\ref{tab.tc}, the values of \tc become $181$\,K in
$Im\bar{3}m$-\hhhs, and $34$\,K in $P\bar{1}$-\hhs.
The anharmonicity reduces \tc by
$21$\,K ($12$\,\%) in the $Im\bar{3}m$
structure~\cite{anharm_comment}, and $10$\,K (29\,\%) in the $P\bar{1}$
structure.
%%The value of \tc including the anharmonicity 
%%agrees well with the experimentally observed transition temperature~\cite{sh3_nat}.
%The effect of the anharmonicity is already discussed by
%Ref.\,\onlinecite{mauri_prl} based on the constant DOS ME theory.
%They include not only the shift of the phonon frequency but also the
%distortion of the phonon polarization vector and insist that the
%latter also has large contribution to the reduction of \tc.
%In our calculation, the polarization is assumed unchanged since we
%completely ignore the mode-mode coupling due to the phonon scattering,
%and obtain the reduction of \tc about $12$\,\% with the anharmonic
%phonon dispersion.
%Since the reduction ratio of \tc in our calculation is smaller than
%that in Ref.\,\onlinecite{mauri_prl} ($19$\,\% at $250$\,GPa), 
%there might be another shift of \tc by considering the change of the
%polarization vector.
Here, the change in \tc by the anharmonicity is the
same order as the shift by both the ZPR and the feedback effect in the
self-consistent calculation. 
This result confirms that 
%the ZPR and the self-consistency of the normal
%self-energy are also important factors for \tc, as well as the anharmonicity. 
the anharmonicity is also important factors in the calculation of \tc, 
as well as the ZPR and the self-consistency of the normal self-energy. 

%Here, the magnitude of the shift of \tc by the anharmonicity
%is the same order of the shift by the ZPR.
%Also from this point of view, we can confirm the importance of the ZPR.
%In addition, in the case of the \hhhs phase, 
%it is important to notice that the constant DOS
%approximation and self-consistency lead to the larger changes of \tc
%than the anharmonicity does.
%It reveals that although the anharmonicity considerably reduce \tc, 
%the treatment of the strongly energy dependent DOS is more crucial for
%the calculation of \tc with the vHs near the Fermi level far from the
%critical pressure.

\begin{table}
 \centering
 \caption{Calculated \tc for $Im\bar{3}m$-\hhhs
 at 250\,GPa and $P\bar{1}$-\hhs at 140\,GPa with different
 methods. 
 %\tc are calculated with 1024 Matsubara frequencies for the
 %$Im\bar{3}m$ structure
 %and with 2048 Matsubara frequencies for the $P\bar{1}$ structure.
 %The Coulomb interaction is included by the RPA screening.
 The first row shows \tc with the constant DOS approximation.
 One-shot and SC denote the one-shot calculation and self-consistent
 calculation for the normal Green's function. 
 ZPR and AH denote the zero-point renormalization for the electron dispersion,
 and the anharmonicity for the phonon frequency, respectively.}
 \label{tab.tc}
\begin{tabular}{c@{\ \ \ \ \ \ } c@{\ \ \ }c@{\ \ \ }}
 \toprule
            & $Im\bar{3}m$-\hhhs & $P\bar{1}$-\hhs \\ 
 \midrule
  const.\,DOS       & 225\,K  & 56\,K \\
  One-shot          & 168\,K  & 66\,K \\ 
  SC	            & 193\,K  & 63\,K \\
 SC $+$ ZPR         & 202\,K  & 44\,K\\
 SC $+$ ZPR  $+$ AH & 181\,K  & 34\,K \\
 \bottomrule
\end{tabular}
\end{table}   
%!TEX root = paper2.tex

\section{Vertex correction}
\label{sec.vertex}

In the ME theory, 
the vertex function is assumed to be dominated by the lowest 
order vertex, which is equal to one, 
and higher order contributions are neglected. 
In order to obtain the quantitative criterion for the justification 
of the ME theory, 
we evaluate the lowest order vertex correction~\cite{schrieffer}. 
Here the lowest order vertex correction 
$\Gamma^{ (1) \lambda}_{j\bm{p+q},l\bm{p}} (i\omega_n + i\omega_m
 ,i\omega_n)$
is given by
\begin{align}
\label{eq.vertex_full}
  &\ \ \ g^{j{\bm p} + {\bm q}, l {\bm p} }_{\lambda} ({\bm q})
 \Gamma^{ (1) \lambda}_{j\bm{p+q},l\bm{p}} (i\omega_n + i\omega_m
 ,i\omega_n) \nonumber \\
%%%
 &= \frac{-1}{N \beta} \sum_{\bm{k}, n'} \sum_{\lambda', j', l'}
 g^{l' {\bm k} + {\bm q}, j' {\bm k}}_{\lambda} (\bm{q})
 G_{l' \bm{k}} (i\omega_{n'})  G_{j' \bm{k+q}}
 (i\omega_{n'}+i\omega_m) \nonumber \\
%%%
 &\times
 g^{j {\bm p} + {\bm q}, j' {\bm k} + {\bm q} }_{\lambda'} (\bm{p-k})
  D_{\bm{p-k}\lambda'}(i\omega_n-i\omega_{n'})
 g^{l' {\bm k}, l {\bm p} }_{\lambda'} (\bm{k-p}).
\end{align}
If the ME theory is justified, 
at least the lowest order correction should be small compared with one.

%The direct evaluation of Eq.\,{\eqref{eq.vertex_full}} is demanding task 
%due to the complicated momenta, band indexes, 
The lowest order vertex correction has a complicated dependence on 
momenta, band indices, frequencies, and phonon modes.
To simplify the evaluation, we here employ several approximations for
Eq.\,\eqref{eq.vertex_full} as follows:
First, phonons are treated as a single Einstein phonon, which does not have 
any momentum dependence. 
Then the phonon Green's function is replaced by
\begin{align}
 D(i\omega_m) = -\frac{2 \langle \omega \rangle}{
 \omega_m^2 + \langle \omega \rangle^2},
\end{align} 
where $\langle \omega \rangle$ is the averaged phonon frequency.
In addition, we ignore the momentum and band index dependence of 
the electron-phonon matrix element,
and replace it with the averaged electron-phonon matrix
element $\langle g \rangle$.
In this case, we could express the averaged
electron-phonon matrix element $\langle g \rangle$ and averaged phonon 
frequency $\langle \omega \rangle$ in terms of the electron-phonon coupling $\lambda$
through the following relation:
\begin{align}
 \lambda = \frac{2 N(0) \langle g \rangle^2}{\langle \omega \rangle}.
\end{align}
With these substitution, the vertex correction only has the 
$\bm{q}$-momentum and 
the Matsubara frequencies dependence, and can be written by
\begin{align}
 \label{eq.vertex}
  \Gamma^{(1)}_{\bm{q}} (i\omega_n + i\omega_m
 ,i\omega_n)
%%%
 &= \frac{\lambda}{N(0)} \frac{1}{N \beta} \sum_{\bm{k}, n'} \sum_{j', l'}
%%%
% D(i\omega_n-i\omega_{n'}) \nonumber \\
 \frac{\langle \omega \rangle^2}{\omega_{n -n'}^2 + \langle \omega \rangle^2} \nonumber \\
 &\times G_{l' \bm{k}} (i\omega_{n'})  G_{j' \bm{k+q}}
 (i\omega_{n'} + i\omega_m).
\end{align}
In Eq.\,\eqref{eq.vertex}, the averaged electron-phonon matrix
element $\langle g \rangle$ does not appear explicitly but has a
contribution only through $\lambda$.
Therefore, we do not need to evaluate $\langle g \rangle$ explicitly.
Finally, the electron Green's function is replaced by the
non-interacting one.

In the practical calculation,
we focus on diagonal elements of the vertex correction in terms of the
Matsubara frequency dependence by setting 
$\omega_m$ as $0$
and evaluate Eq.\,\eqref{eq.vertex} at $\omega_n = \pi/\beta$.
In the case of $Im\bar{3}m$-\hhhs ($P\bar{1}$-\hhs), 
we consider 15 (25) bands and use $100\times100\times100$ ($48\times48\times32$) 
$\bm{k}$-mesh for the sum in the r.h.s.~of Eq.\,\eqref{eq.vertex}.
Since the phonon Green's function works as a convergence factor for 
the Matsubara frequency sum, 
$25$ ($100$) Matsubara frequencies are enough to obtain converged results. 
Here, temperature is set as $200$\,K ($50$\,K) for $Im\bar{3}m$-\hhhs
($P\bar{1}$-\hhs). 
For the phonon Green's function, the $\omega_{\mathrm ln}$ defined by Eq.\,\eqref{eq.omega_ln} is used as 
the averaged phonon frequency $\langle \omega \rangle$. 
The ZPR and anharmonicity are neglected in this calculation.

By using Eq.\,\eqref{eq.vertex} with 
$20\times20\times20$ ($12\times12\times8$) $\bm{q}$-mesh 
for $Im\bar{3}m$-\hhhs ($P\bar{1}$-\hhs), 
we estimate the lowest order vertex correction.
Here, we use $\Delta\Gamma^{(1)}$ calculated by averaging 
$\Gamma^{(1)}_{\bm{q}} (i\omega_n,i\omega_n)$ over $\bm{q}$ as
a measure of the vertex correction,
and $\Delta\Gamma^{(1)}$ is estimated to be $-0.22$ ($-0.12$).
%In the calculation of \tc, such size of the vertex correction causes 
%a shift of \tc by $30$\,K ($16$\,\%) for \hhhs-$Im\bar{3}m$ and 
%$12$\,K ($18$\,\%) for \hhs-$P\bar{1}$, respectively.  
%These changes of \tc's are calculated by using the ME theory 
%with replacing the square of the electron-phonon matrix element with 
%$(1+\Delta\Gamma^{(1)}) |g^{jl}_{\lambda} ({\bm q})|^2$.
%The Coulomb interaction is also enhanced by the same factor like
%$(1+\Delta\Gamma^{(1)}) \tilde{V}^c(\bm{q}, i\omega_m)$.
%Although these shifts of \tc are as large as those by the ZPR or
%anharmonicity, 
%it might be small enough to insist that the ME theory is still available
%in sulfur hydrides.
We then calculate \tc by replacing the square of the electron-phonon
matrix element and the screened Coulomb interaction with 
$(1+\Delta\Gamma^{(1)}) |g^{jl}_{\lambda} ({\bm q})|^2$ and 
$(1+\Delta\Gamma^{(1)}) \tilde{V}^c(\bm{q}, i\omega_m)$, respectively~\cite{vertex_comment}. 
We found that the vertex correction changes \tc by $-34$\,K ($-18$\,\%)
for $Im\bar{3}m$-\hhhs and $-13$\,K ($-20$\,\%) for $P\bar{1}$-\hhs.
We see that the impact of the vertex correction on \tc is similar to
that of ZPR and anharmonicity.

It was shown for a simple model system 
that the lowest order vertex correction is negative in the
static limit of $\omega_m \rightarrow 0$ with finite $\bm{q}$~\cite{pietronero_vertex}.  
Our results are consistent with these previous studies. 
On the other hand, it was also reported that the vertex correction becomes
positive and contributes to the enhancement of \tc in the dynamical limit
($\bm{q} = 0$ with finite $\omega_m$). 
Since these static and dynamical regions in the $\bm{q}$-$\omega_m$ plane 
are approximately separated by a line
of $v_{\mathrm F} |\bm{q}| \approx \omega_m$ where $v_{\mathrm F}$ is the
Fermi velocity, 
the net contribution of the vertex correction %is related
%with the ratio of the volumes of the static and dynamical regions. 
depends on the energy scale of phonons and $v_{\mathrm F}|\bm{q}|$.
%In this sense, 
Our results give the lower bound of \tc corrected
by the inclusion of the vertex function in that we take the limit of 
$\omega_m \rightarrow 0$.  
%%In addition, with vHs, since the dynamical region can be expanded through the
%%reduction of the Fermi velocity, 
%%the vertex correction tends to be positive~\cite{emmanuele_vhs}. 
%%Therefore, it might be a fascinating open question to examine the material dependence of
%the vertex correction with and without vHs in sulfur hydrides. 
%The presence of the vHs, around which $v_{\mathrm F}$ becomes small, 
%could make the dynamical contribution relevant~\cite{emmanuele_vhs}. 
%It is hence an interesting future issue to examine the material dependence 
%of the vertex correction with and without vHs in sulfur hydrides. 
Thus especially for \hhhs having small $v_{\mathrm{F}}$ around vHs, 
the effect of the vertex correction may be overestimated in the present 
calculation, 
since the dynamical contribution can be relevant~\cite{emmanuele_vhs}.
Further study for the vertex function based on a more sophisticated
treatment~\cite{takada_gisc_95} is also an interesting future problem. 

\section{summary} \label{sec.summary}

%In this study, we perform the fully non-empirical calculations of the
%superconducting
%transition temperature for SH$_2$ and SH$_3$ based on the
%self-consistent Migdal-Eliashberg theory, where the energy dependence of
%DOS is strictly included.
%In addition, we can safely include the effect of the Coulomb repulsion
%from first principles.
%By comparing the \tc's by ME theory energy dependent DOS with the
%results assuming the constant DOS, 
%it is confirmed that the constant DOS approximation breaks down in the
%SH$_3$\,-\,$Im\bar{3}m$ structure, where the vHs falls just below the Fermi level, 
%and it results in the reduction of \tc by $57$\,K.
%The zero point renormalization is also included and it causes the
%visible shifts of \tc by $10$\,K.
%It has the same order of the effect of the anharmonicity.
%Although the inclusion of these physical effects 
%leads to changes of \tc of order of ten kelvin, 
%the energy dependence of DOS have the most huge effect in the case of
%the SH$_3$ phases.
%This result denotes that one needs careful considerations 
%about the treatment of superconductivity
%%of the shape of DOS 
%when DOS has narrow peaks around the Fermi level.

One of the most characteristic features in the electronic structure 
of \hhhs under high pressures is the existence of the vHs 
around the Fermi level, 
which is absent for the low \tc phases of \hhs. 
While it has been known that it is crucial to take account of 
the energy dependence of the vHs for accurate estimate of \tc, 
in the previous {\it ab initio} calculation based on the ME theory, 
the constant DOS approximation has been employed.

%In this study, we examine the effect of vHs on \tc of sulfur
%hydrides from first-principles. 
In this study, we performed a self-consistent ME analysis in which 
we explicitly considered the electronic structure over $40$\,eV 
around the Fermi level. 
%In order to understand the role of vHs in superconductivity of sulfur hydrides, 
%we explicitly consider the
%electronic structure over $40$\,eV around the Fermi level 
%within the framework of the ME theory.  
Since \tc's of sulfur hydrides are extremely high,  
%by considering the energy dependence of the DOS, 
%the retardation effect in the Coulomb repulsion can be directly 
%treated without introducing the pseudo Coulomb potential 
%if a sufficiently large number of Matsubara frequencies is employed in
%the Eliashberg equation. 
with a reasonably large number of Matsubara frequencies (up to $\sim 1000$) 
in the Eliashberg equation, 
the retardation effect of the Coulomb interaction could be directly treated. 
By calculating \tc of \hhhs (\hhs), in which vHs are present (absent)
near the Fermi level,
we showed that the constant DOS approximation employed so far seriously
overestimates (underestimates) \tc by $\sim 60$\,K ($\sim 10$\,K). 
In addition, we discussed how the self-energy due to the strong
electron-phonon coupling affects the vHs and \tc, especially focusing
on (1) the feedback effect in the self-consistent calculation of the
self-energy, (2) the effect of the ZPR, and (3) the effect of the
changes in the phonon frequencies due to the strong anharmonicity. 
We showed that the effect of (1)-(3) on \tc is about $10$-$30$\,K for both
\hhhs and \hhs, and eventually \tc is estimated to be $181$\,K for
\hhhs, and $34$\,K for \hhs. 
These results explain the pressure dependence of \tc observed in the
experiment if it is considered that high- (low-)\tc superconductivity
under pressures higher (lower) than $\sim 150$\,GPa is attributed to
that of \hhhs (\hhs). 
Finally, we evaluated the lowest order vertex correction 
%and confirm that the ME theory is good approximation in the case of
%sulfur hydrides.
and we found that its impact on \tc is as large as that of ZPR and anharmonicity.

\begin{acknowledgements}
We thank Mitsuaki Kawamura, Masatoshi Imada,
 Yohei Yamaji, and Yasutami Takada for fruitful discussions.
We also would like to thank Emmanuele Cappelluti for enlightening 
 comments about the vertex correction. 
This work are financially supported by JST, PRESTO and JSPS KAKENHI Grant Numbers
 15K20940 (R.Ak.) and 15H03696 (T.K. and R.Ar.).
This work is partially supported by Tokodai Institute for Element 
 Strategy (TIES) funded by MEXT Elements Strategy Initiative to Form 
 Core Research Center and by the Computational Material Science Initiative (CMSI).
\end{acknowledgements}

\appendix
\section{Density functional perturbation theory} \label{sec.appendix}

In solids, the phonon frequencies are calculated by the following equation:
\begin{eqnarray}
 \label{eq.dynamical_eigen}
 \sum_{\nu \kappa'} D_{\mu\kappa, \nu\kappa'}(\bm{q})
  e^{\nu}_{\kappa'} (\bm{q}\lambda) 
  = \omega^2_{\bm{q} \lambda } e^{\mu}_{\kappa}(\bm{q}\lambda)
\end{eqnarray}
with a momentum ${\bm q}$, ion index $\kappa$, and displacement
direction $\mu, \nu = \{x,y,z\}$. This equation is an eigenvalue problem
for $3n\times3n$ matrix $D(\bm{q})$ with $n$ being the number of atoms.
Therefore the square root of the eigenvalue $\omega_{\bm{q}\lambda}$ and the
polarization vector $e^{\mu}_{\kappa}(\bm{q}\lambda)$ have mode
index $\lambda$, which runs $1 ... 3n$. The matrix $D(\bm{q})$ is
called the dynamical matrix and given by
\begin{eqnarray}
 D_{\mu\kappa, \nu\kappa'}(\bm{q}) = \frac{1}{\sqrt{M_\kappa
  M_{\kappa'}} }
  \Phi_{\mu\kappa, \nu\kappa'}(\bm{q})
\end{eqnarray}
where $M_\kappa$ is the mass of the $\kappa$-th ion, 
%$\bm{r}(l)$ is position of the $l$-th unit cell, 
$\Phi_{\mu\kappa, \nu\kappa'}(\bm{q})$
is the interatomic force constant 
\begin{eqnarray}
 \Phi_{\mu\kappa, \nu\kappa'}(\bm{q})
  = \frac{1}{N}\frac{\partial^2 E(\{\bm{R}^0_\kappa\})}{\partial u_{\mu
  \kappa}^*(\bm{q}) \partial u_{\nu \kappa'}(\bm{q}) }
\end{eqnarray}
with the Born-Oppenheimer energy surface $E(\{\bm{R}^0_\kappa\})$, the
number of ${\bm q}$-points $N$, the
equilibrium positions of the ions $\{\bm{R}^0_\kappa\}$, and the
displacement of the ions $u$.

%In the case with the Hamiltonian depending on sets of the parameters $\{\lambda\}$,
%derivatives of the Born-Oppenheimer energy surface $E_{\lambda}$ around
%ground state configuration are written as
%%Derivatives of the Born-Oppenheimer energy surface can be evaluated with
%%Hellmann-Feynman theorem. 
Derivatives of the Born-Oppenheimer energy surface can be written as
\begin{align}
% \label{eq.th_HF}
%  \frac{\partial^2 E_{\lambda}}{\partial \lambda_i \partial \lambda_j }
%  =& \int  \frac{\partial^2 V_{ie}( {\bm r}) }{\partial \lambda_i
%  \partial \lambda_j} 
%  n ( {\bm r} ) \diff^3 r  \nonumber \\
%& + \int  \frac{\partial V_{\lambda}( {\bm r}) }{\partial \lambda_i}
%  \frac{ \partial n( {\bm r} )}{ \partial \lambda_j } \diff^3 r  \nonumber \\
%  &+   \frac{\partial^2 U_{ii}}{\partial \lambda_i \partial \lambda_j } 
 \label{eq.th_HF}
  \frac{\partial^2 E(\{\bm{R}^0_\kappa\})
 }{\partial u^*_{\mu\kappa}(\bm{q}) \partial u_{\nu\kappa'}(\bm{q}) }
  =& \int  \frac{\partial^2 V_{ie}( {\bm r}) }{\partial u^*_{\mu\kappa}(\bm{q})
  \partial u_{\nu\kappa'}(\bm{q})} 
  n ( {\bm r} ) \diff^3 r  \nonumber \\
& + \int  \frac{\partial V_{ie} ( {\bm r}) }{\partial u^*_{\mu\kappa}(\bm{q})}
  \frac{ \partial n( {\bm r} )}{ \partial u_{\nu\kappa'}(\bm{q}) } \diff^3 r  \nonumber \\
  &+   \frac{\partial^2 U_{ii}}{\partial u^*_{\mu\kappa}(\bm{q}) \partial u_{\nu\kappa'}(\bm{q}) } 
\end{align}
with the electron density $n (\bm{r})$, the ionic potential
$V_{ie}(\bm{r})$ and the ion-ion interaction energy $U_{ii}$. This
formulation can be used for the calculation of the dynamical matrix.

To evaluate Eq.\,(\ref{eq.th_HF}), one needs the response of the electron
density to the ionic displacement. With the Kohn-Sham orbital $\psi_i$, 
the electron density response
 $ \partial n( {\bm r} )/ \partial u_{\mu\kappa}(\bm{q}) $ is given by
\begin{eqnarray}
 \label{eq.density_response}
  \frac{\partial n( {\bm r} ) }{ \partial u_{\mu\kappa}(\bm{q})}  = 4
  {\rm Re} \sum_{i; {\rm occ}} \psi_i^*({\bm r})  \frac{\partial
  \psi_i({\bm r}) }{ \partial u_{\mu\kappa}(\bm{q})},
\end{eqnarray}
where the derivative of the wave function can be written as
\begin{align}
 \label{eq.wf_response}
   (H_{\mathrm{KS}} - \epsilon_i) \left| \frac{\partial \psi_i }{ \partial u_{\mu\kappa}(\bm{q})} \right\rangle 
 = -\left( \frac{\partial  V_{KS} }{ \partial u_{\mu\kappa}(\bm{q})}  -
  \frac{\partial  \epsilon_i }{ \partial u_{\mu\kappa}(\bm{q})}  \right) | \psi_i \rangle.
\end{align}
Here $H_{\mathrm{KS}}$, $V_{\mathrm{KS}}$, and $\epsilon_i$ denote the Hamiltonian, self-consistent potential,
and eigen energy of the Kohn-Sham system respectively.
The derivative of the Kohn-Sham potential also depends on the electron
density response:
\begin{align}
 \label{eq.ks_response}
  \frac{\partial V_{\mathrm{KS}}(\bm{r}) }{ \partial u_{\mu\kappa}(\bm{q})} &= \frac{\partial
   V_{ie}( {\bm r} ) }{ \partial u_{\mu\kappa}(\bm{q})} \nonumber \\
  +& \int  \frac{1}{|{\bm r}- {\bm r}'|} \frac{\partial   n({\bm r}') }{ \partial u_{\mu\kappa}(\bm{q})} \diff^3 r'
  + \frac{\diff V_{xc} }{\diff n}  \frac{\partial   n({\bm r}) }{ \partial u_{\mu\kappa}(\bm{q})}
\end{align}
with the exchange-correlational potential $V_{xc}$. 
%%% to arita 削除しなかった
One can obtain the electronic density response, the Kohn-Sham
wave function response, and the modulation of the potential
simultaneously with solving
Eqs.\,(\ref{eq.density_response}\,-\,\ref{eq.ks_response}) 
as a set of equations. These scheme is known as the Sternheimer method~\cite{dfpt}.
%%% to arita this calculation を this scheme に 変更
Also, the electron-phonon matrix element can be
evaluated through this scheme since it also determined as follows:
\begin{eqnarray}
 \label{eq.elph_matrix}
 g^{i\bm{p} + {\bm q}, j \bm{p} }_{\lambda} ({\bm q}) 
 &=& \sum_{\kappa \mu}\sqrt{\frac{\hbar}{2 M_{\kappa} \omega_{{\bm
 q}\lambda}}} \nonumber \\
 &\times& \langle \psi_{i \bm{p} + {\bm q}} |  \nabla_{l\kappa \mu} V_{\mathrm{KS}} 
 | \psi_{j \bm{p}} \rangle e_{\kappa}^{\mu} ( {\bm q} \lambda),
\end{eqnarray} 
where $\nabla_{l\kappa \mu}$ denotes the partial derivative with respect
to the $\kappa$-th ion position in the $l$-th unit cell for the $\mu$ direction.
This is the formulation of density functional perturbation theory~\cite{dfpt}.
Here, one should notice that the calculated dynamical matrix and electron-phonon matrix element are statically renormalized quantities.

In practical calculation, in order to avoid the singular behavior in
the l.h.s.~of Eq.\,(\ref{eq.wf_response}), one introduces a projection
operator. If Eq.\,(\ref{eq.wf_response}) is formally solved, the derivative of
the wave function is written as
\begin{eqnarray}
  \left| \frac{\partial \psi_i }{ \partial u_{\mu\kappa}(\bm{q})} \right\rangle 
  = \sum_{j \neq i} 
  \frac{\langle \psi_j | \partial  V_{\mathrm{KS}} / \partial u_{\mu\kappa}(\bm{q}) | \psi_i \rangle
  }{\epsilon_i - \epsilon_j} | \psi_j \rangle.
\end{eqnarray}
With this formulation, the electron density response is given by 
\begin{align}
 \label{eq.density_response_2}
  \frac{\partial n( {\bm r} ) }{ \partial u_{\mu\kappa}(\bm{q})}  
  = 4 {\rm Re} &\sum_{i; {\rm occ}} \sum_{j \neq i} 
  \psi_i^*({\bm r})  \psi_j(\bm{r}) \nonumber \\
  &\times \frac{\langle \psi_j | \partial  V_{\mathrm{KS}} / \partial u_{\mu\kappa}(\bm{q}) | \psi_i \rangle
  }{\epsilon_i - \epsilon_j}.
\end{align}
%Although direct calculation of Eq.\,(\ref{eq.density_response_2}) is not
%better solution because of the cumbersome summation over the unoccupied
%stated, this equation tells us useful information. 
%Due to the
%cancellation, summation of the index $j$ in the r.h.s. of
%Eq.\,(\ref{eq.density_response_2}) can be restricted to the unoccupied
%orbitals and the density response only depends on the change of
%the wave function in the unoccupied manifold. 
Eq.\,(\ref{eq.density_response_2}) does not have a convenient form to
solve directly because of the summation over the unoccupied states.
However, due to the cancellation of the contribution involving occupied states, 
the summation over the index $k$ in the r.h.s.~of Eq.\,(\ref{eq.density_response_2})
can be restricted to the unoccupied orbitals and the density response 
only depends on the change of the wave function in the unoccupied manifold.
Therefore, one can evaluate the electron density
response with restriction of Eq.\,(\ref{eq.wf_response}) on the unoccupied
manifold. It is achieved by following equation:
\begin{align}
 \label{eq.stern}
 (H_{\mathrm{KS}} + \alpha P_{\rm occ} - &\epsilon_i) 
 \left| \frac{\partial \psi_i }{ \partial u_{\mu\kappa}(\bm{q})} \right\rangle
 \nonumber \\
 &= -P_{\rm unocc} \frac{\partial  V_{\mathrm{KS}} }{ \partial u_{\mu\kappa}(\bm{q})} | \psi_i \rangle,
\end{align}
where $P_{\rm occ}$ and $P_{\rm unocc}$ are projection to the occupied and
unoccupied spaces, and $\alpha$ is a constant introduced in order to avoid
the singularity of the operator $H_{\mathrm{KS}} - \epsilon_i$. (For the selection
of $\alpha$ and more practical implementation, see e.g.,
Ref.\,\cite{dfpt}.)
One can show that the solution of Eq.\,(\ref{eq.stern}) is equivalent to that of
Eq.\,(\ref{eq.wf_response}) on the unoccupied manifold.

%Also, screened Coulomb interaction is calculated with electronic
%structure obtained from QUANTUM ESPRESSO.
%In this study, screening of electron-electron interaction is
%considered within random phase approximation and dynamical screening is
%ignored. 
%Dynamical structure of the screened Coulomb interaction may cause additional enhancement of \tc through the
%plasmon effect. 
%Inclusion of frequency dependence of the screening is future work. 
%We calculate ZPR based on the Allen-Heine-Cardona (AHC) theory~\cite{gonze_11}
%implemented in ABINIT.


\begin{thebibliography}{99}
\bibitem{w_el_comment} For $Im\bar{3}m$-\hhhs ($P\bar{1}$-\hhs), 
	 \tc's are $217$\,K ($54$\,K) with $\omega_{\mathrm{el}} = 10$\,eV 
	 and $229$\,K ($57$\,K) with $\omega_{\mathrm{el}} = 30$\,eV.
\bibitem{zpr_comment} In the calculation of the ZPR, we use harmonic
	 phonon frequencies. On the other hand, frequencies of
	 bond-stretching phonons which have strong coupling with
	 electrons are hardened due to the anharmonicity. Therefore, the
	 effect of the ZPR might be mitigated by the anharmonicity.
\bibitem{anharm_comment} In Ref.\,\cite{mauri_prl}, it is
	 reported that \tc decreases by $36$\,K ($19$\,\%) with
	 anharmonic phonons in the \hhhs-$Im\bar{3}m$ structure at $250$\,GPa.
	 This reduction is slightly larger than that in our
	 calculation. 
	 Such difference can be attributed to the modulation of the
	 phonon polarization vector, 
	 which is included in the calculation of
	 Ref.\,\cite{mauri_prl} but ignored in ours.
\end{thebibliography}
\end{document}